\documentclass[sigconf, authorversion]{acmart}

\setcopyright{cc}
\setcctype{by}
\acmDOI{10.1145/3832783.3834374}
\acmYear{2026}
\copyrightyear{2026}
\acmISBN{979-8-4007-2882-2/2026/10}
\acmConference[ASE '26]{Proceedings of the 41st IEEE/ACM International Conference on Automated Software Engineering}{October 12--16, 2026}{Munich, Germany}
\acmBooktitle{Proceedings of the 41st IEEE/ACM International Conference on Automated Software Engineering (ASE '26), October 12--16, 2026, Munich, Germany}
\acmSubmissionID{ase26main-p944-p}
\received{2026-03-26}
\received[accepted]{2026-06-18}

\usepackage{hhline}

\usepackage{lipsum}
\usepackage{comment}
\usepackage{amsmath,amsfonts}
\usepackage{graphicx}
\usepackage{textcomp}

\usepackage{booktabs}
\usepackage{colortbl}
\usepackage{tabularx}
\usepackage{adjustbox}
\usepackage{multicol}
\usepackage{multirow}
\usepackage{balance}
\usepackage{xspace}

\usepackage{url}

\usepackage{minted}

\usepackage{listings}
\usepackage{float}
\usepackage{subcaption}

\usepackage{algorithm}
\usepackage{algorithmicx}
\usepackage{algpseudocode}
\usepackage{amsmath}
\floatname{algorithm}{Algorithm}

\definecolor{mypink}{rgb}{1,0.9,0.9}
\definecolor{myblue}{rgb}{0.9,0.9,1}
\definecolor{mygreen}{rgb}{0.9,1,0.9}
\definecolor{myyellow}{rgb}{1,1,0.8}

\usepackage{csquotes}

\usepackage[shortlabels]{enumitem}
\setlist[itemize]{align=parleft,left=0pt..1em}
\setlist[enumerate]{align=parleft,left=0pt..1.5em}

\definecolor{mypink}{rgb}{1,0.9,0.9}
\definecolor{myblue}{rgb}{0.9,0.9,1}
\definecolor{mygreen}{rgb}{0.9,1,0.9}
\definecolor{myyellow}{rgb}{1,1,0.8}

\usepackage[dvipsnames]{xcolor}
\usepackage{pifont}
\definecolor{tiffany}{HTML}{0abab5}

\usepackage{booktabs}
\usepackage{multirow}
\usepackage{graphicx}
\usepackage{amsmath}

\usepackage{tcolorbox}
\tcbset{
    colback=white,
    colframe=black,
    arc=0pt, 
    boxrule=0.5pt,
    top=1pt,
    left=2pt,
    right=2pt,
    bottom=1pt,
    toprule=0.5pt,
    titlerule=0.5pt,
    bottomrule=0.5pt,
    leftrule=0.5pt,
    rightrule=0.5pt}

\newcommand{\rev}[1]{#1} 

\newcommand{\mycomment}[2]{}    

\newcommand{\tool}{\textit{WebCQ}\xspace}
\newcommand{\marg}{\textit{MARG}\xspace}
\newcommand{\webexplor}{\textit{WebExplor}\xspace}
\newcommand{\withoutcomm}{\text{w/o comm.}\xspace}
\newcommand{\withoutqtran}{\text{w/o QTRAN}\xspace}

\begin{document}

\title{WebCQ: Cooperative Multi-agent Deep Reinforcement Learning for Scalable Web GUI Testing}

    
\author{Yujia Fan}
\orcid{0009-0001-8722-3367}
\affiliation{%
  \department{Research Institute of Trustworthy Autonomous Systems, Department of Computer Science and Engineering}
  \institution{Southern University of Science and Technology}
  \city{Shenzhen}
  \country{China}
}
\email{12431253@mail.sustech.edu.cn}

\author{Sinan Wang}
\orcid{0000-0002-8557-4681}
\affiliation{%
  \department{Research Institute of Trustworthy Autonomous Systems, Department of Computer Science and Engineering}
  \institution{Southern University of Science and Technology}
  \city{Shenzhen}
  \country{China}
}
\email{wangsn@mail.sustech.edu.cn}

\author{Zebang Fei}
\orcid{0009-0004-3096-4019}
\affiliation{%
  \department{Research Institute of Trustworthy Autonomous Systems, Department of Computer Science and Engineering}
  \institution{Southern University of Science and Technology}
  \city{Shenzhen}
  \country{China}
}
\email{12110608@mail.sustech.edu.cn}

\author{Yao Qin}
\orcid{0009-0001-6548-4400}
\affiliation{%
  \department{Research Institute of Trustworthy Autonomous Systems, Department of Computer Science and Engineering}
  \institution{Southern University of Science and Technology}
  \city{Shenzhen}
  \country{China}
}
\email{12112016@mail.sustech.edu.cn}

\author{Huaxuan Li}
\orcid{0009-0006-1555-6353}
\affiliation{%
  \department{Research Institute of Trustworthy Autonomous Systems, Department of Computer Science and Engineering}
  \institution{Southern University of Science and Technology}
  \city{Shenzhen}
  \country{China}
}
\email{12112045@mail.sustech.edu.cn}

\author{Yepang Liu}
\orcid{0000-0001-8147-8126}
\authornote{Yepang Liu is the corresponding author.}
\affiliation{%
  \department{Research Institute of Trustworthy Autonomous Systems, Department of Computer Science and Engineering}
  \institution{Southern University of Science and Technology}
  \city{Shenzhen}
  \country{China}
}
\email{liuyp1@sustech.edu.cn}

\begin{abstract}

Multi-agent reinforcement learning (MARL)-based techniques have shown promise for GUI testing.
However, as the complexity of modern GUI software increases, existing MARL-based approaches (e.g., \marg and \textit{Fastbot}) struggle to scale due to the inherent limitations of their underlying tabular reinforcement learning algorithms.
This limits their applicability to large-scale commercial GUI software, especially web applications with vast state spaces and many interactive elements.
To fill this gap, we propose \tool, a novel MARL-based approach for scalable web GUI testing.
\tool incorporates QTRAN for multi-agent coordination and a lightweight synchronization mechanism, allowing it to work under asynchronous web testing scenarios.
It extracts semantic and exploration features for each UI event to form an action vector.
This vector is concatenated with the current state vector and fed into the policy network, enabling DQN-based decision making within a dynamic action space.
We evaluated \tool on eight large-scale commercial websites.
Under the same time budget and agent count, \tool explored 33.3\% more states and executed 42.2\% more unique actions than \marg, while triggering more failures on six of the eight websites under test.
It also demonstrated strong scalability, maintaining higher action throughput during 20-hour experiments, and achieving greater performance improvements as the number of agents increased.
These results show that \tool overcomes key limitations of existing MARL-based approaches, providing a scalable and effective solution to enhance modern web GUI testing.

\end{abstract}

\ccsdesc[300]{Software and its engineering~Software testing and debugging}
\ccsdesc[100]{Computing methodologies~Multi-agent systems}

\keywords{Web GUI Testing, Multi-agent Reinforcement Learning}

\maketitle

\section{Introduction}

GUI-based software systems have grown increasingly complex, particularly in large-scale web applications, making automated and scalable testing techniques more important~\cite{halani2021critical, banerjee2013graphical, lilarge}.
Among the techniques, end-to-end GUI testing explores applications by generating UI event sequences to simulate open-ended user interactions, aiming to achieve high functional coverage and expose bugs.

Recent studies have demonstrated the potential of \textit{reinforcement learning} (RL) in GUI testing~\cite{zheng2021automatic,fan2023comprehensive,mariani2011autoblacktest,mariani2014automatic,koroglu2018qbe,pan2020reinforcement}.
For example, \textit{WebExplor}~\cite{zheng2021automatic} and \textit{QExplore}~\cite{sherin2023qexplore} employ the tabular RL algorithm Q-learning~\cite{watkins1992q} for web exploration.
Despite their differences in algorithmic configurations, they both significantly outperform testing approaches based on traditional techniques (e.g., model-based~\cite{biagiola2019diversity} or search-based~\cite{biagiola2017search} ones).
Nonetheless, tabular RL algorithms can easily suffer from the \textbf{state explosion} problem, where the state-action policies (i.e., Q-tables) can quickly grow unbounded as the number of states increases, limiting their practical usages in testing complex web applications.
To mitigate this problem, researchers have explored the alternative use of \textit{deep reinforcement learning}~(DRL).
DRL agents leverage deep neural networks (DNNs) to capture high-dimensional or continuous state representations, thereby avoiding unbounded Q-table growth. 
For example, Romdhana et al.~\cite{romdhana2022deep} proposed an Android GUI testing framework and evaluated multiple exploration strategies, including policy-based DRL algorithms such as DDPG, TD3, and SAC, as well as baselines like Q-learning and random exploration.
They show that SAC achieves the best performance.
\textit{DQT}~\cite{lan2024deeply}, a DQN-based Android testing approach, employs graph embedding techniques to effectively identify similarities among application states and actions.
This accelerates the discovery of high-value actions, improves instruction coverage, and uncovers more unique failures.

The above approaches employ single-agent paradigms, whose performance is inherently bounded by the limitations of isolated learning.
More recently, researchers have explored the use of \textit{multi-agent reinforcement learning} (MARL) to explore complex GUI environments via parallelized execution and experience sharing.
For example, \textit{Fastbot}~\cite{cai2020fastbot} runs multiple Q-learning agents to collectively build navigation models for Android applications.
Fan et al.~\cite{fan2024can} show that running multiple RL agents independently leads to significant redundancy in their explored states.
They propose \marg, a framework that incorporates two novel communication schemes to enable effective experience sharing among concurrent RL-based testing agents.
Compared to independently running RL agents, \marg~can explore significantly more web states and detect more unique failures.
However, these multi-agent approaches mostly rely on tabular RL algorithms, leading to severe \textbf{communication overhead} between parallelized agents.
In addition, their locator-centric action definitions cannot adequately capture \textbf{action semantics}, i.e., the intended functionality behind an action (e.g., submitting a form). 
Adequately capturing such semantics is important because it not only preserves the meaning of critical user tasks but also enables more effective validity analysis of dynamic interactions (e.g., clicking elements with unstable XPaths).

This paper proposes \tool, a novel MARL-based approach that addresses the limitations of existing work.
\tool targets complex web applications, aiming to achieve more effective exploration and improved coordination among parallel testing agents.
We formulate multi-agent web GUI testing task as a Dec-POMDP problem~\cite{oliehoek2016concise, hansen2004dynamic}, and adopt the \textit{centralized training and decentralized execution} (CTDE) paradigm~\cite{zhang2021multi} to design our workflow.
In centralized training, the communication schemes proposed by {\marg} are designed for tabular RL algorithms, making them inefficient for complex websites due to excessive communication overhead.
To achieve better coordination, \tool employs QTRAN~\cite{son2019qtran}, a more advanced MARL algorithm.
Since QTRAN assumes synchronized agents, \tool introduces a \textbf{lightweight synchronization mechanism} that supports asynchronous exploration while ensuring that sufficient experience is collected for joint training. 
By avoiding unnecessary waiting, this mechanism can improve overall efficiency.
Furthermore, a hybrid reward function is introduced to encourage agents to explore diverse states, avoid repeating actions, and focus on long-term exploration.
In decentralized execution, each agent preserves the structural information of a state while encoding richer information into the action vectors.
Instead of using only UI element locators, \tool's action representation focuses on maintaining \textbf{semantic and exploration-relevant information}. 
To handle the dynamic action space, where predefining a fixed action set for the policy network is infeasible, \tool concatenates the state vector with each action vector and feeds the combined representation into the policy network for action selection.

To demonstrate \tool's effectiveness, we compared it with {\webexplor} (originally designed as a single-agent approach but adapted in our work to run with multiple agents for a fair comparison), the state-of-the-art MARL-based GUI exploration approach {\marg}, 
as well as two ablation variants, both driven by DRL agents: one without communication, and one without using the QTRAN algorithm.
These approaches were all evaluated on eight diversified and large-scale commercial websites. 
Our experiments show that, with the same time budget and the same number of agents, \tool significantly outperforms all other approaches in terms of the number of explored states, executed unique actions, and detected web failures.
Furthermore, \tool also exhibits superior scalability, achieving higher hourly executed actions during long-term (20 hours) explorations and greater performance gains as the number of agents increases.

In summary, our work makes the following contributions:

\begin{itemize}
\item \rev{To the best of our knowledge, we are the first to formulate the multi-agent web GUI testing task as a Dec-POMDP problem. Building on this formulation, we adapt QTRAN to the asynchronous web GUI testing setting, thereby enabling centralized training and cooperation among testing agents.}
\item We propose a novel MARL-based web GUI testing approach, \tool, and demonstrate that it achieves state-of-the-art performance in terms of exploration efficiency, failure-triggering effectiveness, and scalability.
\item To facilitate future research and industry adoption, we have released the source code of \tool and all experimental data in: \rev{\url{https://doi.org/10.5281/zenodo.21679869}}.
\end{itemize}

\section{Preliminaries}

\subsection{Reinforcement Learning}
\label{ssec:dqn}

Reinforcement learning (RL) studies how an autonomous agent can learn to make sequential decisions by interacting with an environment to maximize cumulative rewards~\cite{padakandla2021survey, sutton2018reinforcement}.
A key branch of RL is \textit{tabular RL algorithm} (e.g., Q-learning~\cite{watkins1992q}), which estimates a Q-value function $Q(s,a)$ to approximate the expected future reward for taking a specific action $a$ in a given state $s$. 
However, tabular methods struggle to handle high-dimensional and continuous state spaces.
\rev{To overcome this limitation, the Deep Q-Network~(DQN)~\cite{mnih2015human} leverages a DNN to approximate the Q-value function $Q(s, a; \theta)$, where $\theta$ denotes the network weights.}
DQN addresses DNN instability in RL through two pivotal mechanisms:
\begin{enumerate}
\item \textbf{Experience replay}: During exploration, state transitions with reward $(s,a,s',r)$ will be stored in a replay buffer, from which mini-batches will be sampled for training. 
\item \textbf{Target network}: A separate network with weights $\theta^-$ periodically copies the main network's weights $\theta$. 
It decouples the target value from the current prediction to stabilize training.
\end{enumerate}
They allow the DNN loss to be computed as the mean square error between the target and the prediction over the mini-batch:
\begin{equation}
\label{eq:vdn}
\mathcal{L}(\theta)=\mathbb{E}_{(s,a,s',r)}\Bigl[(r+\gamma\max_{a'}Q(s',a';\theta^-)-Q(s,a;\theta))^2\Bigr]
\end{equation}

\subsection{Cooperative MARL and QTRAN}
\label{ssec:qtran}

\begin{figure}[t]
\centering
\includegraphics[width=\columnwidth]{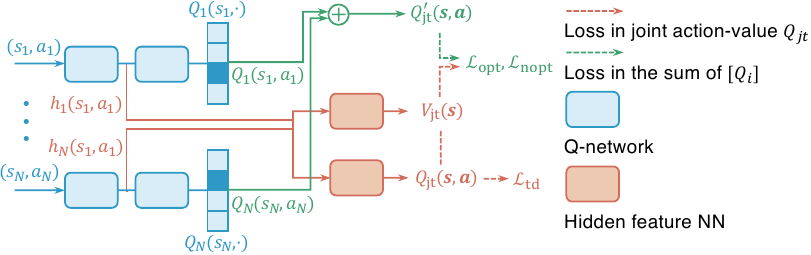}
\vspace{-2em}
\caption{QTRAN's architecture~\cite{son2019qtran}.}
\label{fig:qtran}
\end{figure}

\begin{figure*}[t]
\centering
\includegraphics[width=0.85\textwidth]{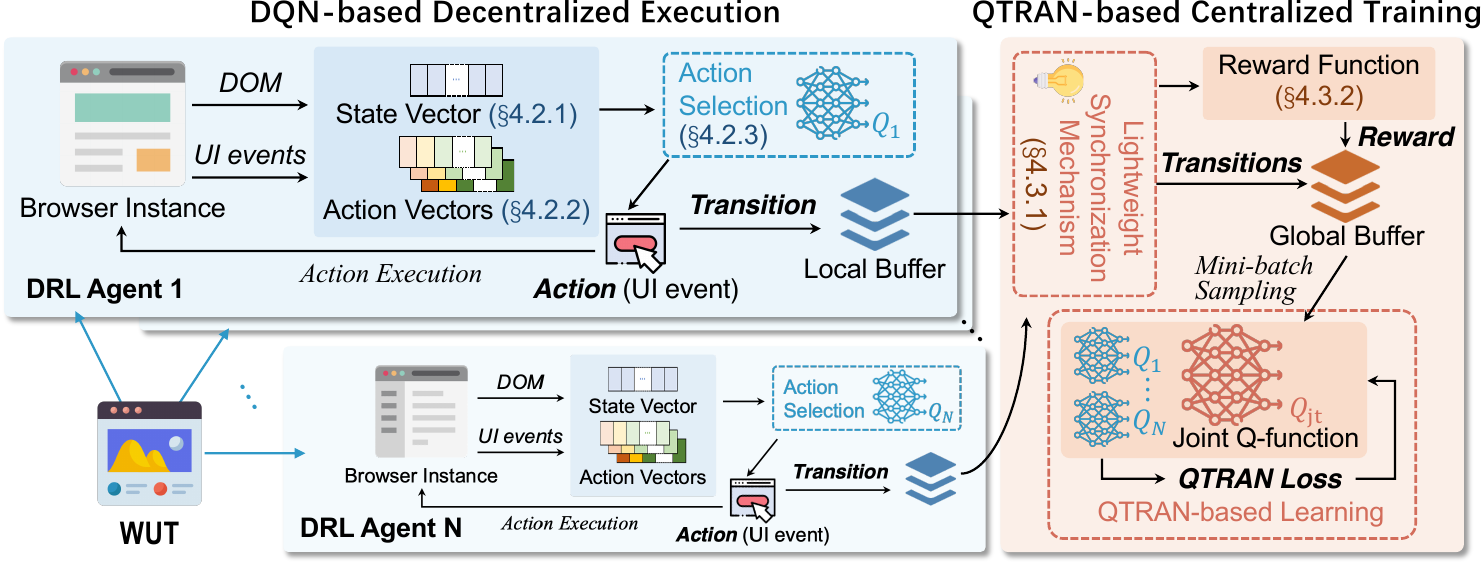}
\caption{An overview of \tool.}
\label{fig:overview}
\end{figure*}

In cooperative MARL tasks with CTDE paradigm~\cite{zhang2021multi}, agents select actions based on partial observations but learn strategies through a global policy.
This requires each agent to maintain its own Q-function $Q_i(s_i,a_i)$ while contributing jointly to the optimization of a global Q-function $Q_\textrm{jt}(\boldsymbol{s},\boldsymbol{a})$, where $\boldsymbol{s}$ denotes the \textit{joint state} and $\boldsymbol{a}$ denotes the \textit{joint action}.
However, the challenge lies in how to effectively decompose the global Q-value into individual agent contributions during training.
QTRAN (Figure~\ref{fig:qtran}) introduces a state-value function $V_\textrm{jt}(\boldsymbol{s})$ to account for the discrepancy between the joint Q-function $Q_\textrm{jt}$ and the 
\rev{sum of individual utilities $Q_i$}:
\begin{equation}
V_\textrm{jt}(\boldsymbol{s})=\max_{\boldsymbol{a}'}Q_\textrm{jt}(\boldsymbol{s},\boldsymbol{a}')-\sum_{i=1}^N\max_{a_i'}Q_i(s_i,a_i')
\end{equation}
This allows the joint model to capture complex interactions among agents without being restricted by additivity or monotonicity assumptions.
Then, QTRAN's loss function can be defined as the weighted sum of three components:
\begin{align}\begin{split}
\label{eq:loss}
\mathcal{L}_\textrm{td} &= \mathbb{E}\Bigl[(r+\gamma\max_{\boldsymbol{a}'}Q_\textrm{jt}(\boldsymbol{s}',\boldsymbol{a}';{\boldsymbol\theta}^-)-Q_\textrm{jt}(\boldsymbol{s}, \boldsymbol{a}))^2\Bigr] \\
\mathcal{L}_\textrm{opt} &= \mathbb{E}\Bigl[(\max_{\boldsymbol{a}'}Q'_\textrm{jt}(\boldsymbol{s},\boldsymbol{a}')-\max_{\boldsymbol{a}'}\hat{Q}_\textrm{jt}(\boldsymbol{s},\boldsymbol{a}')+V_\textrm{jt}(\boldsymbol{s}))^2\Bigr] \\
\mathcal{L}_\textrm{nopt} &= \mathbb{E}\Bigl[(\min[Q'_\textrm{jt}(\boldsymbol{s}, \boldsymbol{a})-\hat{Q}_\textrm{jt}(\boldsymbol{s}, \boldsymbol{a})+V_\textrm{jt}(\boldsymbol{s}), 0])^2\Bigr]
\end{split}\end{align}
where $Q_\textrm{jt}'=\sum_{i=1}^N{Q_i}$ and $\hat{Q}_\textrm{jt}$ indicates that computing the losses $\mathcal{L}_\textrm{opt}$ and $\mathcal{L}_\textrm{nopt}$ will not update $Q_\textrm{jt}$.
The $\mathcal{L}_{td}$ employs the standard temporal difference error to update the $Q_\textrm{jt}$, which is similar to DQN (Equation \ref{eq:vdn}).
$\mathcal{L}_\textrm{opt}$ and $\mathcal{L}_\textrm{nopt}$ describe the constraints on $Q_\textrm{jt}$ for taking the optimal joint action and other suboptimal actions, respectively.
Then the individual Q-functions $[Q_i]$ are learned through learning $Q_\textrm{jt}$, $Q_\textrm{jt}'$ and $V_\textrm{jt}$.

\section{Problem Formulation}
\label{sec:formulation}

Given a WUT (website under test), we adopt a multi-agent testing setting where multiple agents interact with their own browser instances to explore the website in parallel.
Each agent makes decisions based on the webpage it sees, while a global objective is to achieve comprehensive exploration and trigger potential failures.

This process can be formulated as a decentralized partially observable Markov decision process (Dec-POMDP)~\cite{oliehoek2016concise, hansen2004dynamic}, defined as $\langle N, \boldsymbol{S}, \{A_i\}_{i\in \{1,...,N\}},P,\{R_i\}_{i\in \{1,...,N\}},\gamma,\{O_i\}_{i\in \{1,...,N\}}\rangle$,
where $N$ is the number of agents, $\mathcal{S}$ denotes the global state space, $A_i$ and $O_i$ represent the action space and observation space of agent $i$, respectively, $P$ is the state transition function, $R_i$ is the local reward function, and $\gamma$ is the discount factor. 
In Dec-POMDP, each agent cannot observe the global state and each agent has the same reward function~\cite{zhang2021multi}.
At each timestep, agent $i$ has a local state $s_i \in O_i$, denoting its observation for notational consistency with QTRAN.

\textit{Definition 1 }(\textbf{Local State}):
For agent $i$, a local state $s_i$ represents the current webpage observed through its browser instance.
A webpage is represented by its HTML document, which encodes both content and structural information. 
Existing work such as \webexplor converts HTML pages into tag sequences and measures similarity to identify equivalent states~\cite{zheng2021automatic}.
However, such methods can incur substantial computational overhead for complex webpages~\cite{liujudge, fan2023comprehensive}.
To address this limitation, we will propose a more efficient abstract representation for local states (see Section~\ref{ssse:state}). 

\textit{Definition 2} (\textbf{Action}):
An action $a\in A$ represents a UI event in the WUT, which corresponds to an interaction with a DOM element (e.g., clicking a button or entering text into an input field).
Since the set of available UI events varies across webpages, the action space in web testing is dynamically determined by the current state.

\textit{Definition 3} (\textbf{Reward}): 
Each agent $i$ receives a local reward $r_i$ after performing an action, reflecting its contribution to exploration. 
\rev{A global reward, defined as $r = \sum_{i=1}^{N} r_i$, is used for training in the cooperative multi-agent setting.}
Existing work typically designs reward functions based on state novelty~\cite{romdhana2022deep, lan2024deeply} and action diversity~\cite{zheng2021automatic, lan2024deeply} to encourage exploration. 
Our hybrid reward function also takes these principles into account (see Section~\ref{sssec:reward}).

\section{Approach}
\label{sec:approach}

\subsection{Overview}
\label{ssec:overview}

CTDE is widely adopted to address the coordination and partial observability in Dec-POMDPs~\cite{foerster2017stabilising, gupta2017cooperative}.
In cooperative tasks, CTDE enables coordinated learning by allowing a centralized learner to access joint states, actions, and rewards during training, while each agent executes with its own policy based on local states.
Given these advantages, QTRAN, an MARL method under the CTDE paradigm, becomes appropriate for the multi-agent web testing task, where effective coordination among agents is crucial.

We propose \tool, a cooperative multi-agent web testing approach that follows the CTDE paradigm.
As illustrated in Figure~\ref{fig:overview}, the framework decomposes the testing process into two components:
(1) DQN-based Decentralized Execution and (2) QTRAN-based Centralized Training.
Given a WUT, each agent launches an independent browser instance to access the webpage.
The agent constructs the state and action vectors, selects an action using its Q-network, and executes it to trigger a transition stored in the local buffer.
The transitions are iteratively collected and sent to the central module for centralized training. 
To determine when centralized training will be performed, we propose a lightweight synchronization mechanism to coordinate the asynchronous agents. 
Using the collected data of all agents, the transitions and the computed global reward are stored in the global buffer.
Mini-batches are then sampled from the global buffer to update the Q-networks through QTRAN, optimizing both the joint and individual Q-functions.
In the following sections, we describe the two components in detail.

\subsection{DQN-based Decentralized Execution}
\label{ssec:de}

\begin{figure}[t]
    \centering
    \includegraphics[width=\linewidth]{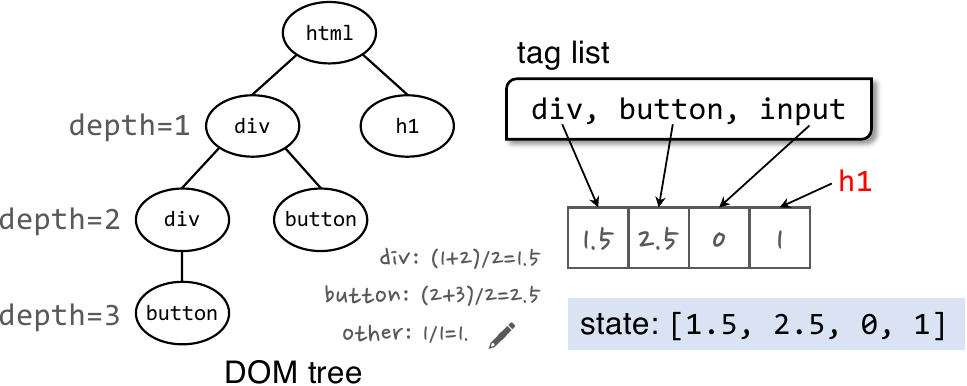}
    \caption{An example of tag-depth-based state vector.}
    \label{fig:state}
\end{figure}

\subsubsection{State Vector}
\label{ssse:state}
Existing work such as \webexplor represents a state as a sequence of HTML tags and relies on sequence similarity for state matching, which incurs substantial computational overhead. 
One-hot encodings of activities used in Android GUI testing~\cite{romdhana2022deep} are also unsuitable for web testing, where identifiers like URLs are often dynamic \rev{and cannot be known in advance.}
\rev{Besides, in single-page applications (SPAs), the same URL may correspond to different DOM states~\cite{mesbah2012crawling}.}

\rev{Different from existing approaches, \tool captures DOM structural information for state representation by encoding HTML tags and their corresponding DOM depths.}
As illustrated in Figure~\ref{fig:state}, given a predefined list of common HTML tags, we compute the average DOM depth~\cite{w3schoolsW3Schoolscom} of elements for each tag (zero for absent tags) to construct the state vector, with an extra dimension for tags outside the list.
This method produces vectors of a fixed number of dimensions, enabling efficient state matching and compatibility with neural network models.

\subsubsection{Action Vector}
\begin{figure}[!t]
     \centering
     \includegraphics[width=\linewidth]{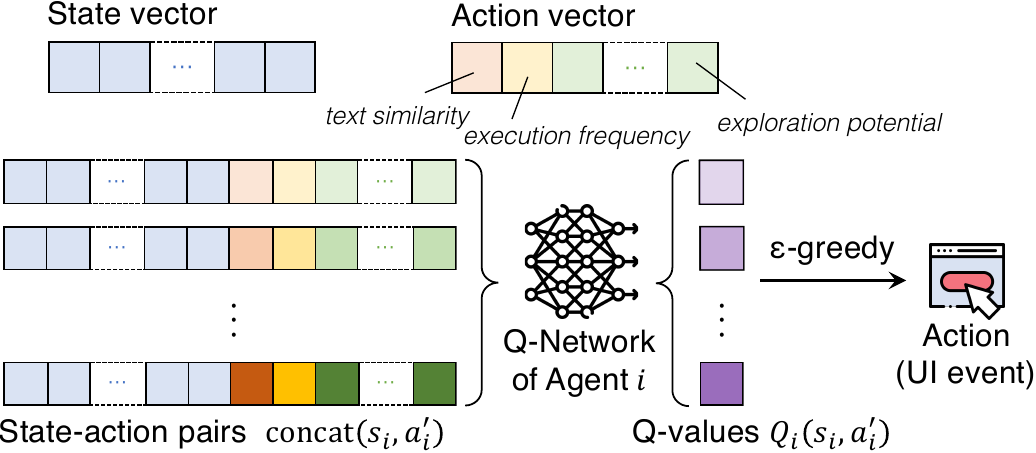}
     \caption{Action selection strategy of the testing agent.}
     \label{fig:action_selection}
\end{figure}

While the state vector captures the structural features of the current page, distinguishing among candidate actions fundamentally depends on their specific features. 
These include the textual semantics of actions, which often indicate their potential interaction effects~\cite{zhao2024dinodroid}, as well as exploration-relevant features, such as actions that are rarely executed or likely to lead to states with richer exploration opportunities.
Accordingly, each candidate action on the current state is transformed into a vector composed of three key components (Figure~\ref{fig:action_selection}):

\begin{itemize}
    \item \textit{Text similarity of current actions.}
Buttons containing positive keywords, such as \texttt{accept}, \texttt{confirm}, \texttt{submit}, \texttt{ok}, \texttt{yes}, or \texttt{next}, are more likely to trigger transitions to new pages and support efficient exploration.
To quantify this, the button text and predefined positive keywords are encoded using a pre-trained \textit{GloVe} model~\cite{pennington2014glove}, and cosine similarity is computed for each keyword.
The highest similarity score serves as a feature representing how closely the current action aligns with typical positive actions.

\item \textit{Execution frequency of current action.}
The execution count of each UI event is recorded and incorporated into the corresponding action vector.
This allows the agent to be aware of repetitive actions, which in turn are discouraged via execution frequency term in the reward function (Section~\ref{sssec:reward}), guiding the agent toward more diversified exploration.

\item \textit{Exploration potential of the child state.}
In addition to execution frequency, the exploration potential of the child state (i.e., the webpage after executing a given action) also reflects the action’s quality.
We represent this using a 10-dimensional vector, where each entry counts how many actions in the direct child state have been executed a given number of times.
For example, if a child state contains ten actions, four never executed,
six executed once, and one executed more than eight times, then the vector would be $[4,6,0,0,0,0,0,0,0,1]$.
For an unexplored action whose child state is unknown, the vector is filled with zeros.

\end{itemize}

The action vector for each UI event evolves dynamically during exploration. UI events with positive keywords, few prior interactions, and high exploration potential tend to have higher Q-values, while the repeating UI events gradually receive lower Q-values as their click counts increase.
This adaptive mechanism allows the agent to avoid redundant behaviors and prioritize under-explored states and actions, thereby enhancing the efficiency of exploration within the time budget.

\subsubsection{Action Selection Strategy}

For each agent, we use a DQN to parameterize the Q-function, following QTRAN's value-based formulation.
However, in web testing, the available UI events vary across states, resulting in a dynamic and state-dependent action space.
This characteristic is inconsistent with the standard DQN, which assumes a fixed action space and predicts Q-values over a predefined set of actions.
To address this gap, for a given state, each candidate action is combined with the state vector and fed as input to the Q-function, as illustrated in Figure~\ref{fig:action_selection}.
At each decision step, agent~$i$ concatenates the state vector with each available action vector to form a set of state–action pairs, which are then used as inputs to the policy network~$f$ to evaluate their $Q$-values:
\begin{equation}
\label{eq:concat}
Q_i(s_i, a^{\prime}_{i}) = f(\texttt{concat}(s_i, a^{\prime}_{i})) 
\end{equation}
Based on the computed Q-values, the agent selects an action using an $\varepsilon$-greedy strategy~\cite{dann2022guarantees}:
\begin{align}
\label{eq:epsilon-greedy}
a_i^* = \left\{
\begin{array}{ll}
    \mathop{\mathrm{argmax}}\limits_{a^{\prime}_{i}} {Q_i(s_i, a^{\prime}_{i})} & \text{with\ probability} \ 1\mathit{-}\varepsilon, \\
    \text{random action} & \text{with probability} \  \varepsilon
\end{array} \right.
\end{align}

\rev{It is worth noting that $a_i^{\prime}$ is an action vector used for RL training and action selection; for execution, it is mapped to the corresponding UI action, specified by its action type and the unique XPath locator of its target element.}

\subsection{QTRAN-based Centralized Training}
\label{ssec:qtran_po}

\subsubsection{Adaptation}
In the centralized training phase, we employ QTRAN to optimize each agent's policy. 
QTRAN was originally designed for synchronous tasks in shared environments~\cite{vinyals2017starcraftiinewchallenge,cui2019uav}, where agents observe different parts of a common global state.
However, in multi-agent web testing, each agent operates an independent browser with its own webpage context. 
Therefore, agents’ observations are not drawn from a shared spatial-temporal environment, and action execution across agents is inherently asynchronous.
To fill this gap, we introduce two strategies to adapt QTRAN to work in the asynchronous web testing scenario:

\ding{172}\,\textbf{Joint state representation.}
In our task, each agent can only observe its own local state, and no explicit global state exists.
Meanwhile, agents’ local states may overlap because they observe pages from the same WUT, either at the same time or at different times.
To represent the joint state of all agents, we construct $\boldsymbol{s}$ by concatenating the local state vectors $s_i$ of each agent: $\boldsymbol{s} = (s_1; s_2; \dots; s_N)$.
This joint state preserves the unique perspectives of individual agents and provides the joint Q-function with a comprehensive view of the WUT for learning coordinated policies.

\ding{173}\,\textbf{Lightweight synchronization mechanism.}
A straightforward approach is to synchronize agents after each step; however, this is inefficient as faster agents must wait for the slowest one.
\tool instead enables asynchronous execution, allowing agents to explore independently without step-level blocking.
Once each agent has collected at least one new transition, their latest experiences are aggregated to update the QTRAN network.
This design preserves asynchronous exploration while ensuring that training proceeds only when sufficient new samples are available.

\subsubsection{Reward Function}
\label{sssec:reward}
Following our problem formulation (Section~\ref{sec:formulation}), the global reward for centralized training is computed based on each agent's individual reward, that is, $r = \sum_{i=1}^{N} r_i$.
Designing an effective reward function is crucial to guide agents toward meaningful exploration.
Existing studies often design rewards based on state novelty~\cite{romdhana2022deep, lan2024deeply} or action diversity~\cite{zheng2021automatic, lan2024deeply}. 
Building on this idea, we define a hybrid reward that encourages agents to explore diverse states, avoid repeating actions, and focus on long-term exploration:
\begin{equation}
r = (R_{\text{sim}} + R_{\text{exec}}) \times (1 + R_{\text{time}})
\end{equation}

\noindent The details of each component are as follows:
\begin{itemize}
\item\textit{Page similarity}  $R_\text{sim}$:
To encourage exploration of diverse pages, we first identify the most similar previously visited page $p^*$ of the current page $p$ using the APIMiner metric~\cite{chen2024apiminer}. 
The $R_\text{sim}$ is then evaluated piecewise:
\begin{equation}
R_{\text{sim}} = 
\begin{cases}
    r_\text{high}, & \text{similarity}(p, p^*) < \sigma_\text{lower} \\
    r_\text{mid}, & \sigma_\text{lower} \leq \text{similarity}(p, p^*) < \sigma_\text{upper} \\
    r_\text{low}, & \text{similarity}(p, p^*) \geq \sigma_\text{upper},\ N(p)=0 \\
    \frac{1}{N(p)}, & \text{otherwise}
\end{cases}
\label{eq:state_reward}
\end{equation}
where $N(p)$ denotes the visit count of webpage $p$.

\item\textit{Execution frequency} $R_{\text{exec}}$: 
Since similar pages often share identical UI events, repeated executions usually yield redundant exploration. 
To mitigate this, $R_{\text{exec}}$ assigns a bonus to new actions and provides a diminishing reward for repeated executions:
\begin{equation}
R_{\text{exec}} = 
\begin{cases}
    2, & N(e) = 1 \\
    \frac{1}{N(e)}, & N(e) > 1
\end{cases}
\end{equation}
where $N(e)$ is the total execution count of event $e$ of all agents.

\item\textit{Temporal compensation} $R_{\text{time}}$:
During long-term exploration, rewards can become sparse: many states are already explored, leading to diminishing rewards. 
Following \textit{DQT}~\cite{lan2024deeply}, we use a time-based reward term $R_{\text{time}}$ that increases with execution time:
\begin{equation}
R_{\text{time}} = \frac{t}{t_{\text{total}}}.
\end{equation}
where $t_{\text{total}}$ denotes time budget.

\end{itemize}

\begin{algorithm}[t]
\caption{\tool's workflow}
\label{alg:qtran}
\begin{algorithmic}[1]
\State \textbf{Input:} Number of agents $N$, Mini-batch size $M$
\State \textbf{Initialize:} For each agent $i$, initialize $Q_i$ and done signal $d_i \gets \texttt{false}$; Initialize $Q_{\text{jt}}$, replay buffer $\mathcal{B}$, optimization lock, and centralized training signal $ct \gets \texttt{false}$


\For{each agent $i$ \textbf{in parallel}}
\label{alg1:Asynchronous}
\Repeat
    \State \Call{AsynchronousExecution}{$i$}
\Until{time budget exhausts}
\EndFor


\Procedure{AsynchronousExecution}{$i$}
    \Repeat
        \State \textbf{wait()} \Comment{Pause if centralized training in progress}
    \Until{$ct = \texttt{false}$}
    \label{alg1:until}

    \State $s_i \gets$ get current state embedding
    \label{alg1:get_si}
    \State $a_i \gets$ select action using Equation (\ref{eq:epsilon-greedy})
    \label{alg1:get_ai}
    \State $s_i' \gets$ execute action $a_i$
    \label{alg1:ae_exec}
    \State $\textit{trans}_i \gets (s_i, a_i, s'_i)$
    \label{alg1:ae_trans}
    \State $d_i \gets \texttt{true}$
    \label{alg1:ae_di}


    \State lock.acquire()                                                       \label{alg1:lock-acquire}
    \If{$\forall i\ d_i = \texttt{true}$ \textbf{and} $ct = \texttt{false}$}    \label{alg1:ct_codi}
        \State $ct \gets \texttt{true}$                                         \label{algo1:ct-true}
        \State \Call{CentralizedTraining}{\null}
        \State $ct \gets \texttt{false}$                                        \label{algo1:ct-false}
    \EndIf
    \State lock.release()

\EndProcedure


\Procedure{CentralizedTraining}{\null}
    \State $\textit{trans} \gets (\boldsymbol{s}, \{a_i\}_N, \boldsymbol{s}')$    \label{algo1:joint-state}
    \Comment{{$\boldsymbol{s}=\texttt{concat}(\{s_i\}_N)$}}
    \State$r\gets \sum_{i=1}^N r_i$ 
    \label{algo1:reward}
    \Comment{Compute $r_i$ using reward function}
    \State Store $\textit{trans}$ and $r$ into $\mathcal{B}$                                                \label{algo1:joint-state-store}


    \If{$|\mathcal{B}| \geq M$}                                                         \label{algo1:batch-size}
        \State Sample mini-batch $\{(\boldsymbol{s}, \boldsymbol{a}, \boldsymbol{s^{\prime}}, r)\}_M \sim \mathcal{B}$                 \label{algo1:compute-loss}
        \State Compute $\mathcal{L}_{\text{td}}, \mathcal{L}_{\text{opt}}, \mathcal{L}_{\text{nopt}}$ using Equation (\ref{eq:loss})
        \State $\mathcal{L}_{\text{total}} \gets \mathcal{L}_{\text{td}} + \mathcal{L}_{\text{opt}} + \mathcal{L}_{\text{nopt}}$
        \State Update $Q_\text{jt}$ and $[Q_i]_N$ based on $\mathcal{L}_\text{total}$   \label{algo1:update-loss}
    \EndIf

    \For{each agent $i$}
        \State $d_i \gets \texttt{false}$                                               \label{algo1:reset-done-flag}
    \EndFor
\EndProcedure

\end{algorithmic}
\end{algorithm}

\subsubsection{Workflow}
Algorithm~\ref{alg:qtran} shows how \tool works.
At a high level, each agent continuously interacts with its own environment in parallel (Line~\ref{alg1:Asynchronous}) until the given time budget is exhausted.
After completing a state transition, the agent records the transition locally and sets its done signal to \texttt{true}~(Lines~\ref{alg1:get_si}-\ref{alg1:ae_di}).
By acquiring the global optimization lock, the agent checks whether all agents have completed at least one transition and whether centralized training is not already in progress~(Lines~\ref{alg1:lock-acquire}-\ref{alg1:ct_codi}).
Using the centralized training signal \textit{ct}, the thread of the last agent to complete its transition triggers the centralized training procedure~(Lines~\ref{algo1:ct-true}-\ref{algo1:ct-false}), while other agents wait for it to finish (Line~\ref{alg1:until}).

Due to the asynchronous execution of agents, some may occasionally perform one additional transition before reacquiring the lock in the next iteration.
Allowing agents to perform one extra step can reduce webpage loading delays between episodes, thereby improving overall efficiency.
Before QTRAN-based learning, the most recent transitions of all agents are merged into a joint transition, and a global reward $r$ is computed based on the agents' transitions. 
The joint transition and the reward are then stored in the replay buffer (Lines~\ref{algo1:joint-state}-\ref{algo1:joint-state-store}).
If the buffer has a sufficient number of samples (Line~\ref{algo1:batch-size}), we compute the QTRAN loss and update both the joint Q-network and each agent’s individual Q-network (Lines~\ref{algo1:compute-loss}-\ref{algo1:update-loss}).
After training, all done signals are reset (Line~\ref{algo1:reset-done-flag}), allowing agents to continue exploration.
This lightweight mechanism avoids blocking faster agents unnecessarily while still ensuring coordinated updates based on recent experiences.

\section{Experimental Setup}
\label{sec:setup}

\begin{table}[t]
    \centering
    \caption{Settings of the compared approaches and \tool.}
    \label{tab:evaluated_approaches}
    \resizebox{\columnwidth}{!}
    {

\begin{tabular}{c|c|c|c|c|c} 
\toprule
\multicolumn{1}{c|}{\textbf{Group}} &
  \textbf{Approach} &
  \textbf{Description} &
  \textbf{$\gamma$} &
  \textbf{$\alpha$} &
  \textbf{$\varepsilon$} \\ 
\midrule
\rowcolor[HTML]{F5F5F5} 
&
  \textit{\webexplor}~\cite{zheng2021automatic} &
  Single Q-learning agent &
  0.95 &
  &
  \\   
\hhline{~---~~}
\rowcolor[HTML]{F5F5F5} 
\multirow{-2}{*}{\begin{tabular}[c]{@{}l@{}}SOTA\\Methods\end{tabular}} &
  \textit{\marg}~\cite{fan2024can} &
  \begin{tabular}[c]{@{}c@{}}Multiple Q-learning agents \\ with communication\end{tabular} &
  0.5 &
  \multirow{-2}{*}{1} &
  \multirow{-2}{*}{0.5} \\  
\midrule
\rowcolor[HTML]{FFF8E1} 
&
  \tool \withoutcomm &
 \begin{tabular}[c]{@{}c@{}}Multiple DRL agents \\ without communication\end{tabular} &
  &
  &
  \\
\hhline{~--~~~}
\rowcolor[HTML]{FFF8E1} 
\multirow{-3}{*}{\begin{tabular}[c]{@{}l@{}}Ablation\\Variants\end{tabular}} &
  \tool \withoutqtran &
  \begin{tabular}[c]{@{}c@{}}Multiple DRL agents \\ without QTRAN\end{tabular} &
  \multirow{-3}{*}{0.5} &
  \multirow{-3}{*}{0.001} &
  \multirow{-3}{*}{\begin{tabular}[c]{@{}c@{}}$\varepsilon_{max}=0.9$\\$\varepsilon_{min}=0.3$\end{tabular}} \\  
\midrule
\rowcolor[HTML]{E3F2FD} 
\begin{tabular}[c]{@{}l@{}}Proposed\\Approach\end{tabular} &
  $\tool$ &
  \begin{tabular}[c]{@{}c@{}}Multiple DRL agents \\ with QTRAN algorithm\end{tabular} &
  0.5 &
  0.001 &
  \begin{tabular}[c]{@{}c@{}}$\varepsilon_{max}=0.9$\\$\varepsilon_{min}=0.3$\end{tabular}\\ 
\bottomrule
\end{tabular}

}
\end{table}

\subsection{Research Questions}

To evaluate the performance of \tool, we conducted a series of experiments to investigate the following research questions:

\begin{itemize}
    \item \textbf{RQ1 (Tool Performance)}: How does {\tool} perform when testing real-world web applications, and how does it compare to the state-of-the-art methods?
    \item \textbf{RQ2 (Ablation Study)}: What is the impact of multi-agent communication and the use of the QTRAN algorithm on {\tool}'s performance?
    \item \textbf{RQ3 (Coordination Overhead)}: Can {\tool} reduce the coordination overhead over parallel agents during the testing period compared to the tabular-RL approach {\marg}?
    \item \textbf{RQ4 (Effect of Agent Numbers)}: How does the number of agents affect the overall performance of {\tool} when testing large-scale websites?
\end{itemize}

\subsection{Compared Approaches}

\subsubsection{State-of-the-Art Methods}
For RQ1, we evaluated $\tool$ against two representative state-of-the-art web testing tools.
Table~\ref{tab:evaluated_approaches} summarizes their key features and major parameters.

{\webexplor} is a representative single-agent, Q-learning-based web testing approach, with a DFA dynamically built for error recovery~\cite{zheng2021automatic}.
We selected {\webexplor} as a baseline since it achieves the state-of-the-art performance among single-agent web testing approaches~\cite{zheng2021automatic}.
To fairly compare with \tool in a multi-agent scenario, we simultaneously run $N$ independent \webexplor processes and merge their testing results, denoted as $\webexplor^N$.

The state-of-the-art MARL-based web GUI testing approach, \marg, implements a distributed Q-learning framework with direct data exchange among asynchronous testing agents~\cite{fan2024can}.
It represents a state by the URL together with the available UI event set, and employs a curiosity-based reward function to encourage exploration.
We selected its distributed Q-learning version (i.e., $\marg_\textit{D}$), which has been shown to be more efficient for large-scale websites.

\subsubsection{Ablation Variants}
\label{sssec:ablation-variants}
For RQ2, we developed two ablation variants of $\tool$ to assess the impact of multi-agent communication and the QTRAN algorithm.
They are also summarized in Table~\ref{tab:evaluated_approaches}, with descriptions highlighting the components that were removed.

The first variant, \tool \withoutcomm, has multiple DRL agents operating independently and in parallel, without any communication.
Each agent uses a DQN, with the same state and action representations, reward, and action selection mechanism as $\tool$.

The second variant, \tool \withoutqtran, allows multiple DRL agents to communicate by sharing experiences. 
In this ablation, QTRAN is replaced with a weaker communication scheme from \marg.
Specifically, when agent $i$ observes a next state $s'$ for which other agents have visited, it updates its own Q-network according to these past experiences, following the rule: 
\begin{align}
\label{eq:dql}
Q_i(s,a) \leftarrow \ \alpha \bigg[ r + \frac{\gamma}{l} 
\sum_{Q_j \in \mathcal{Q}_{-i}(s')} Q_j\Big(s', \mathop{\mathrm{argmax}}_{a'} Q_i(s', a') \Big) \bigg]\notag\\ +(1-\alpha)\ Q_i(s,a)
\end{align}
\noindent Here, $l$ is the number of other agents that have visited $s'$, $\mathcal{Q}_{-i}(s')$ denotes the set of their Q-functions. 
The Q-value $Q_i(s, a)$ for agent $i$ is updated using the received reward $r$, the agent’s current Q-value, and the average estimated future rewards from other agents.

\subsection{Configurations}

For {\webexplor} and \marg's hyper-parameters, we retained the values of discount factor $\gamma$, learning rate $\alpha$ and exploration rate $\varepsilon$ from their original papers.
For \tool and its ablation variants, we set $\gamma=0.5$ as well.
However, considering the different magnitudes of value updates in DRL networks compared to Q-table, we adjusted the learning rate $\alpha$ to 0.001~\cite{schaul2015prioritized}.
To better adapt to the dynamics of DNN training, we adopted a linearly decaying exploration strategy~\cite{zhang2023convergence},
such that $\varepsilon$ decreases linearly over time:
\begin{equation}
\varepsilon(t) = \varepsilon_{\text{max}} - \min\left(\frac{2t}{T_{\text{total}}}, 1\right) \times (\varepsilon_{\text{max}} - \varepsilon_{\text{min}})
\end{equation}
where $\varepsilon_{\text{max}}=0.9$ is the initial maximum exploration rate, $\varepsilon_{\text{min}}=0.3$ is the minimum exploration rate, $t$ is the current testing time and $T_{\text{total}}$ is the testing time budget.
The exploration rate gradually decays during the first half of the testing session and remains at its minimum value during the second half.
In the early phase (first $T_{\text{total}}/2$ time), while the policy network has not yet converged, a high exploration rate ($0.3\le\varepsilon\le0.9$) is maintained to allow extensive sampling, preventing premature convergence to local optima.
As the learning process stabilizes, the exploration rate linearly decreases to its minimum, allowing the agent to gradually rely more on the learned policy.

We set the page similarity component weights to $\omega_{\text{url}}=0.75$, $\omega_{\text{form}}=0.25$, and $\omega_{\text{js}}=0.1$, and the similarity thresholds to $\sigma_\text{lower}=0.7$ and $\sigma_\text{upper}=0.85$.
The similarity reward scores were set to $r_\text{high}=50$, $r_\text{mid}=10$, and $r_\text{low}=2$.
These hyperparameters were tuned through a pilot experiment.

We denote each approach with a superscript $N$ to specify the number of running agents.
For example, $\tool^5$ represents \tool configured with five testing agents.

\subsection{Subject Websites}

\begin{table}[t]
    \centering
    \caption{Our subject websites (WUTs).}
    \label{tab:subjects}
    \vspace{-0.5em}
    \resizebox{0.9\columnwidth}{!}
    {\begin{tabular}{lll}
\toprule
\textbf{Name}   & \textbf{Category}         & \textbf{URL}                      \\
\midrule
EatingWell      & Food and Beverages        & \url{www.eatingwell.com}      \\
ESPN            & Sports                    & \url{www.espn.com}      \\
FoxNews        & Newspapers                    & \url{www.foxnews.com}         \\
GameRant       & Computer and Video Games   & \url{www.gamerant.com}     \\
GameSpot        & Entertainment             & \url{www.gamespot.com}        \\
Gap             & Apparel and Fashion       & \url{www.gap.com}         \\
GitHub          & Software and Development      & \url{www.github.com}   \\
Smadex          & Advertising and Marketing & \url{www.smadex.com} \\

\bottomrule
\end{tabular}
}
\end{table}

Compared to single-agent testing, multi-agent approaches are particularly suited for large-scale websites.
Therefore, we constructed our subject list based on \marg’s benchmark, prioritizing websites with richer state spaces to better reflect this setting.
Websites such as Toppr and Vuestic, which exhibit relatively small state spaces in \marg’s results, were not considered.
In addition, due to the limitation of our embedding model, which currently supports only English~\cite{githubReleaseGlovewikigigaword300}, websites with substantial non-English content (YouTube and IKEA) were excluded.
We then included four additional WUTs randomly sampled from top-ranked websites~\cite{semrush} across different categories.
Table~\ref{tab:subjects} lists the final subject websites.

\subsection{Implementation and Environment}

Similar to \marg's practice, \tool also adopts a client-server architecture.
On the agent side, web testing automation is supported by Selenium-Python, while the centralized controller schedules agents using Python’s built-in threading module.
All policy networks are implemented in PyTorch (version 2.3.1) with default weight initialization.
Specifically, each network is a DenseNet~\cite{huang2018denselyconnectedconvolutionalnetworks} of four fully connected layers, where every layer receives the concatenated outputs of all preceding layers.

We ran all our experiments on a server running Ubuntu 22.04 and Chrome 117.
The server has 32 CPU cores (64 threads), 128GB RAM, and two NVIDIA Quadro RTX 6000 GPUs, and is connected to the Internet with a Gigabit Ethernet.


\section{Experimental Results}

\subsection{RQ1: Tool Performance}

\subsubsection{Method}
\label{ssec:rq1_method}

We compared \tool against two state-of-the-art methods, \webexplor and \marg, all configured with five agents.
We ran each method on each WUT for three hours and repeated six times to mitigate the threat of randomness.
The total experimental time is comparable to that of existing studies~\cite{zheng2021automatic,lan2024deeply,fan2024can,bauersfeld2014user}.

Following prior work~\cite{mesbah2008crawling,fan2024can,wang2024leveraging}, we use the number of \textit{explored states}, \textit{executed unique actions}, and \textit{detected failures} to measure the performance of web exploration approaches.
Notably, we used the definition of ``\textbf{URL with element set}'' to represent the metric ``explored states'', which is consistent with that in {\marg}'s experimental setting.
\rev{The metric ``detected failures''  was computed from the web browser's console logs. Specifically, we retained only SEVERE-level logs, normalized their contents (e.g., by replacing numbers with \texttt{<num>}), and counted the unique normalized logs.}

\subsubsection{Result}
\label{ssec:rq1_res}

\begin{table*}[t]
    \centering
    \caption{Comparisons of state-of-the-art methods (\webexplor and \marg), ablation variants ($\tool$ \withoutcomm and $\tool$ \withoutqtran), and our approach $\tool$, each with five agents.}
    \label{tab:rq1}
    \resizebox{0.92\textwidth}{!}
    {

\renewcommand{\arraystretch}{1.2}
\begin{tabular}{c|cc|rrrrrrrr|r}
\toprule
\textbf{Metric (\#)} &
  \multicolumn{2}{c|}{\textbf{Approach}} &
  EatingWell &
  ESPN &
  FoxNews &
  GameRant &
  GameSpot &
  Gap &
  GitHub &
  Smadex &
  \textbf{Average} \\ \midrule
 &
  \multicolumn{1}{c|}{\cellcolor[HTML]{F5F5F5}} &
  \cellcolor[HTML]{F5F5F5}$\webexplor^5$ &
  \cellcolor[HTML]{F5F5F5}27.0 &
  \cellcolor[HTML]{F5F5F5}52.0 &
  \cellcolor[HTML]{F5F5F5}70.8 &
  \cellcolor[HTML]{F5F5F5}59.0 &
  \cellcolor[HTML]{F5F5F5}18.0 &
  \cellcolor[HTML]{F5F5F5}48.8 &
  \cellcolor[HTML]{F5F5F5}21.7 &
  \cellcolor[HTML]{F5F5F5}112.2 &
  \cellcolor[HTML]{F5F5F5}51.2 \\
 &
  \multicolumn{1}{c|}{\multirow{-2}{*}{\cellcolor[HTML]{F5F5F5}\begin{tabular}[c]{@{}c@{}}SOTA\end{tabular}}} &
  \cellcolor[HTML]{F5F5F5}$\marg^5$ &
  \cellcolor[HTML]{F5F5F5}783.7 &
  \cellcolor[HTML]{F5F5F5}1118.5 &
  \cellcolor[HTML]{F5F5F5}787.8 &
  \cellcolor[HTML]{F5F5F5}939.2 &
  \cellcolor[HTML]{F5F5F5}111.7 &
  \cellcolor[HTML]{F5F5F5}1043.0 &
  \cellcolor[HTML]{F5F5F5}711.3 &
  \cellcolor[HTML]{F5F5F5}\textbf{377.8} &
  \cellcolor[HTML]{F5F5F5}734.1 \\
 &
  \multicolumn{1}{c|}{\cellcolor[HTML]{FFF8E1}} &
  \cellcolor[HTML]{FFF8E1}$\tool^5$ \withoutcomm &
  \cellcolor[HTML]{FFF8E1}871.0 &
  \cellcolor[HTML]{FFF8E1}1021.7 &
  \cellcolor[HTML]{FFF8E1}854.7 &
  \cellcolor[HTML]{FFF8E1}1022.3 &
  \cellcolor[HTML]{FFF8E1}85.5 &
  \cellcolor[HTML]{FFF8E1}1205.0 &
  \cellcolor[HTML]{FFF8E1}818.2 &
  \cellcolor[HTML]{FFF8E1}281.2 &
  \cellcolor[HTML]{FFF8E1}769.9 \\
 &
  \multicolumn{1}{c|}{\multirow{-2}{*}{\cellcolor[HTML]{FFF8E1}\begin{tabular}[c]{@{}c@{}}Ablation\end{tabular}}} &
  \cellcolor[HTML]{FFF8E1}$\tool^5$ \withoutqtran &
  \cellcolor[HTML]{FFF8E1}867.2 &
  \cellcolor[HTML]{FFF8E1}1022.5 &
  \cellcolor[HTML]{FFF8E1}977.0 &
  \cellcolor[HTML]{FFF8E1}1037.3 &
  \cellcolor[HTML]{FFF8E1}99.2 &
  \cellcolor[HTML]{FFF8E1}1214.5 &
  \cellcolor[HTML]{FFF8E1}887.7 &
  \cellcolor[HTML]{FFF8E1}348.5 &
  \cellcolor[HTML]{FFF8E1}806.7 \\
\multirow{-5}{*}{\begin{tabular}[c]{@{}c@{}}Explored\\ States\end{tabular}} &
  \multicolumn{1}{c|}{\cellcolor[HTML]{E3F2FD}\begin{tabular}[c]{@{}c@{}}\end{tabular}} &
  \cellcolor[HTML]{E3F2FD}\textbf{$\tool^5$} &
  \cellcolor[HTML]{E3F2FD}\textbf{1008.5} &
  \cellcolor[HTML]{E3F2FD}\textbf{1229.3} &
  \cellcolor[HTML]{E3F2FD}\textbf{1063.8} &
  \cellcolor[HTML]{E3F2FD}\textbf{1227.2} &
  \cellcolor[HTML]{E3F2FD}\textbf{115.3} &
  \cellcolor[HTML]{E3F2FD}\textbf{1613.7} &
  \cellcolor[HTML]{E3F2FD}\textbf{1215.7} &
  \cellcolor[HTML]{E3F2FD}356.7 &
  \cellcolor[HTML]{E3F2FD}\textbf{978.8} \\ \midrule
 &
  \multicolumn{1}{c|}{\cellcolor[HTML]{F5F5F5}} &
  \cellcolor[HTML]{F5F5F5}$\webexplor^5$ &
  \cellcolor[HTML]{F5F5F5}36.8 &
  \cellcolor[HTML]{F5F5F5}50.3 &
  \cellcolor[HTML]{F5F5F5}105.3 &
  \cellcolor[HTML]{F5F5F5}82.2 &
  \cellcolor[HTML]{F5F5F5}53.7 &
  \cellcolor[HTML]{F5F5F5}57.5 &
  \cellcolor[HTML]{F5F5F5}43.2 &
  \cellcolor[HTML]{F5F5F5}153.7 &
  \cellcolor[HTML]{F5F5F5}72.8 \\
 &
  \multicolumn{1}{c|}{\multirow{-2}{*}{\cellcolor[HTML]{F5F5F5}\begin{tabular}[c]{@{}c@{}}SOTA\end{tabular}}} &
  \cellcolor[HTML]{F5F5F5}$\marg^5$ &
  \cellcolor[HTML]{F5F5F5}1025.3 &
  \cellcolor[HTML]{F5F5F5}1318.0 &
  \cellcolor[HTML]{F5F5F5}970.2 &
  \cellcolor[HTML]{F5F5F5}1034.7 &
  \cellcolor[HTML]{F5F5F5}189.5 &
  \cellcolor[HTML]{F5F5F5}975.0 &
  \cellcolor[HTML]{F5F5F5}895.3 &
  \cellcolor[HTML]{F5F5F5}457.5 &
  \cellcolor[HTML]{F5F5F5}858.2 \\
 &
  \multicolumn{1}{c|}{\cellcolor[HTML]{FFF8E1}} &
  \cellcolor[HTML]{FFF8E1}$\tool^5$ \withoutcomm &
  \cellcolor[HTML]{FFF8E1}1207.0 &
  \cellcolor[HTML]{FFF8E1}1315.8 &
  \cellcolor[HTML]{FFF8E1}1093.0 &
  \cellcolor[HTML]{FFF8E1}1246.2 &
  \cellcolor[HTML]{FFF8E1}197.0 &
  \cellcolor[HTML]{FFF8E1}1398.3 &
  \cellcolor[HTML]{FFF8E1}1084.2 &
  \cellcolor[HTML]{FFF8E1}461.5 &
  \cellcolor[HTML]{FFF8E1}1000.4 \\
 &
  \multicolumn{1}{c|}{\multirow{-2}{*}{\cellcolor[HTML]{FFF8E1}\begin{tabular}[c]{@{}c@{}}Ablation\end{tabular}}} &
  \cellcolor[HTML]{FFF8E1}$\tool^5$ \withoutqtran &
  \cellcolor[HTML]{FFF8E1}1229.8 &
  \cellcolor[HTML]{FFF8E1}1212.3 &
  \cellcolor[HTML]{FFF8E1}1076.7 &
  \cellcolor[HTML]{FFF8E1}1232.3 &
  \cellcolor[HTML]{FFF8E1}208.7 &
  \cellcolor[HTML]{FFF8E1}1360.8 &
  \cellcolor[HTML]{FFF8E1}1198.8 &
  \cellcolor[HTML]{FFF8E1}529.5 &
  \cellcolor[HTML]{FFF8E1}1006.1 \\
\multirow{-5}{*}{\begin{tabular}[c]{@{}c@{}}Executed\\ Unique\\ Actions\end{tabular}} &
  \multicolumn{1}{c|}{\cellcolor[HTML]{E3F2FD}\begin{tabular}[c]{@{}c@{}}\end{tabular}} &
  \cellcolor[HTML]{E3F2FD}\textbf{$\tool^5$} &
  \cellcolor[HTML]{E3F2FD}\textbf{1535.7} &
  \cellcolor[HTML]{E3F2FD}\textbf{1585.8} &
  \cellcolor[HTML]{E3F2FD}\textbf{1159.3} &
  \cellcolor[HTML]{E3F2FD}\textbf{1501.5} &
  \cellcolor[HTML]{E3F2FD}\textbf{246.8} &
  \cellcolor[HTML]{E3F2FD}\textbf{1838.7} &
  \cellcolor[HTML]{E3F2FD}\textbf{1357.5} &
  \cellcolor[HTML]{E3F2FD}\textbf{539.2} &
  \cellcolor[HTML]{E3F2FD}\textbf{1220.6} \\ \midrule
 &
  \multicolumn{1}{c|}{\cellcolor[HTML]{F5F5F5}} &
  \cellcolor[HTML]{F5F5F5}$\webexplor^5$ &
  \cellcolor[HTML]{F5F5F5}17.5 &
  \cellcolor[HTML]{F5F5F5}21.0 &
  \cellcolor[HTML]{F5F5F5}59.7 &
  \cellcolor[HTML]{F5F5F5}12.8 &
  \cellcolor[HTML]{F5F5F5}24.8 &
  \cellcolor[HTML]{F5F5F5}25.3 &
  \cellcolor[HTML]{F5F5F5}11.2 &
  \cellcolor[HTML]{F5F5F5}9.2 &
  \cellcolor[HTML]{F5F5F5}22.7 \\
 &
  \multicolumn{1}{c|}{\multirow{-2}{*}{\cellcolor[HTML]{F5F5F5}\begin{tabular}[c]{@{}c@{}}SOTA\end{tabular}}} &
  \cellcolor[HTML]{F5F5F5}$\marg^5$ &
  \cellcolor[HTML]{F5F5F5}23.2 &
  \cellcolor[HTML]{F5F5F5}45.3 &
  \cellcolor[HTML]{F5F5F5}\textbf{93.7} &
  \cellcolor[HTML]{F5F5F5}47.5 &
  \cellcolor[HTML]{F5F5F5}34.2 &
  \cellcolor[HTML]{F5F5F5}37.2 &
  \cellcolor[HTML]{F5F5F5}32.8 &
  \cellcolor[HTML]{F5F5F5}27.2 &
  \cellcolor[HTML]{F5F5F5}42.6 \\
 &
  \multicolumn{1}{c|}{\cellcolor[HTML]{FFF8E1}} &
  \cellcolor[HTML]{FFF8E1}$\tool^5$ \withoutcomm &
  \cellcolor[HTML]{FFF8E1}23.8 &
  \cellcolor[HTML]{FFF8E1}47.2 &
  \cellcolor[HTML]{FFF8E1}74.0 &
  \cellcolor[HTML]{FFF8E1}45.0 &
  \cellcolor[HTML]{FFF8E1}31.3 &
  \cellcolor[HTML]{FFF8E1}34.2 &
  \cellcolor[HTML]{FFF8E1}\textbf{36.7} &
  \cellcolor[HTML]{FFF8E1}25.2 &
  \cellcolor[HTML]{FFF8E1}39.7 \\
 &
  \multicolumn{1}{c|}{\multirow{-2}{*}{\cellcolor[HTML]{FFF8E1}\begin{tabular}[c]{@{}c@{}}Ablation\end{tabular}}} &
  \cellcolor[HTML]{FFF8E1}$\tool^5$ \withoutqtran &
  \cellcolor[HTML]{FFF8E1}\textbf{29.2} &
  \cellcolor[HTML]{FFF8E1}44.8 &
  \cellcolor[HTML]{FFF8E1}76.7 &
  \cellcolor[HTML]{FFF8E1}47.3 &
  \cellcolor[HTML]{FFF8E1}\textbf{34.3} &
  \cellcolor[HTML]{FFF8E1}36.5 &
  \cellcolor[HTML]{FFF8E1}35.0 &
  \cellcolor[HTML]{FFF8E1}28.7 &
  \cellcolor[HTML]{FFF8E1}41.6 \\
\multirow{-5}{*}{\begin{tabular}[c]{@{}c@{}}Detected\\ Failures\end{tabular}} &
  \multicolumn{1}{c|}{\cellcolor[HTML]{E3F2FD}\begin{tabular}[c]{@{}c@{}}\end{tabular}} &
  \cellcolor[HTML]{E3F2FD}\textbf{$\tool^5$} &
  \cellcolor[HTML]{E3F2FD}27.5 &
  \cellcolor[HTML]{E3F2FD}\textbf{51.3} &
  \cellcolor[HTML]{E3F2FD}73.5 &
  \cellcolor[HTML]{E3F2FD}\textbf{48.8} &
  \cellcolor[HTML]{E3F2FD}33.0 &
  \cellcolor[HTML]{E3F2FD}\textbf{41.3} &
  \cellcolor[HTML]{E3F2FD}34.7 &
  \cellcolor[HTML]{E3F2FD}\textbf{45.3} &
  \cellcolor[HTML]{E3F2FD}\textbf{44.4} \\ \bottomrule
\end{tabular}
}
\end{table*}

As shown in Table~\ref{tab:rq1}, \tool demonstrates a significant improvement in exploration capability.
It achieves the highest number of explored states on seven out of the eight WUTs, with an average of 978.8 explored states, outperforming the state-of-the-art technique \marg by a percentage of $33.3\%(=(978.8-734.1)/734.1)$.
In terms of action diversity, \tool also achieves the most executed unique actions, with an average count of 1220.6, which is $42.2\%(=(1220.6-858.2)/858.2)$ higher than \marg.

Meanwhile, \webexplor, which parses HTML to represent states as tag sequences, suffers from excessive computational overhead in state matching. 
This inefficiency becomes more prominent on large-scale websites with complex DOM structures, and thus, even when aggregating the exploration results of five parallel agents, its performance remains poor on all WUTs.

\begin{table}[t]
\centering
\caption{\rev{Average per-run distribution of failure types reported by \tool on each WUT.}}
\label{tab:failures}
\resizebox{\columnwidth}{!}{
\rev{
\begin{tabular}{lrrrrrr}
\toprule
\textbf{WUT}
& \textbf{Render}
& \textbf{Res.}
& \textbf{API/Async.}
& \textbf{Runtime}
& \textbf{Sec./Auth.}
& \textbf{Others} \\
\midrule
EatingWell & 3.0\%  & \textbf{34.9\%} & 14.6\% & 20.9\% & 20.4\% & 6.3\%  \\
ESPN       & 4.6\%  & \textbf{33.4\%} & 4.7\%  & 26.8\% & 18.3\% & 12.2\% \\
FoxNews    & 3.4\%  & 16.2\% & 5.3\%  & \textbf{42.1\%} & 24.4\% & 8.5\%  \\
GameRant   & 1.1\%  & 10.6\% & 2.0\%  & \textbf{63.2\%} & 19.1\% & 4.0\%  \\
GameSpot   & 1.4\%  & 19.3\% & 25.2\% & \textbf{32.6\%} & 16.1\% & 5.4\%  \\
Gap        & 22.9\% & \textbf{32.7\%} & 20.3\% & 13.8\% & 6.8\%  & 3.5\%  \\
GitHub     & 30.5\% & 13.1\% & 3.6\%  & 13.2\% & \textbf{37.6\%} & 1.9\%  \\
Smadex     & 4.6\%  & 11.6\% & 6.3\%  & \textbf{37.0\%} & 30.7\% & 9.8\%  \\
\midrule
\textbf{Average}
& {8.9\%}
& {21.5\%}
& {10.2\%}
& \textbf{31.2\%}
& {21.7\%}
& {6.4\%} \\
\bottomrule
\end{tabular}
}
}
\end{table}

On most WUTs, \tool outperforms \marg by a clear margin.
However, on GameSpot, \tool achieves slightly higher explored states than \marg, while on Smadex, \marg performs slightly better than \tool.
Notably, both WUTs have relatively fewer states compared to the others.
This reveals that \marg's effectiveness is limited to less complex websites with relatively small state spaces.
In addition, \marg triggers fewer unique actions, suggesting that its increase in state count is due to less interactive webpages, leading to more visited states without enhancing action diversity.

Although \tool is primarily designed to comprehensively explore the WUT and does not include specialized mechanisms for bug triggering, broader state coverage naturally provides more opportunities to expose web failures. 
As shown in Table~\ref{tab:rq1}, \tool detects more failures than \marg on six of the eight WUTs.

\rev{To examine the types of failures reported by \tool, we classified all collected failure logs into six categories: rendering and UI, resource loading, API/asynchronous request, runtime and application logic, security/authentication, and others~\cite{marchetto2007empirical,ocariza2013empirical}.
Table~\ref{tab:failures} shows their distributions.
Across the eight WUTs, an average of 93.6\% of the failures fall into the five major categories.}

\rev{Table~\ref{tab:failures} shows that the distributions of failure categories vary considerably across the WUTs.
Runtime and application logic failures have the highest average proportion (31.2\%) and constitute the dominant category in four of the eight WUTs.
In contrast, resource loading is the dominant category on three of the eight WUTs (i.e., EatingWell, ESPN, and Gap).
For example, \tool recorded \texttt{PhotoCarousel error: image failed to load} for two different product images on two different Gap product pages.
}

\subsubsection{Statistical Analysis}

\begin{table*}[t]
    \centering
    \caption{
    Pairwise statistical comparison of $\tool$ against other approaches (each with five agents) using Mann-Whitney U test. Statistical significance is determined at $p<0.05$, while $A_{12}>0.71$ indicates a large effect.}
    \label{tab:rq1_utest}
    \vspace{-0.8em}
    \resizebox{0.93\textwidth}{!}
    {
\renewcommand{\arraystretch}{1.2}

\begin{tabular}{cccccccccccccccc}
\toprule
\multicolumn{2}{c}{} &
   &
  \multicolumn{2}{c}{\textbf{\begin{tabular}[c]{@{}c@{}}Explored \\ States\end{tabular}}} &
  \multicolumn{2}{c}{\textbf{\begin{tabular}[c]{@{}c@{}}Executed \\ Unique\\ Actions\end{tabular}}} &
  \multicolumn{2}{c}{\textbf{\begin{tabular}[c]{@{}c@{}}Detected \\ Failures\end{tabular}}} &
   &
  \multicolumn{2}{c}{\textbf{\begin{tabular}[c]{@{}c@{}}Explored \\ States\end{tabular}}} &
  \multicolumn{2}{c}{\textbf{\begin{tabular}[c]{@{}c@{}}Executed \\ Unique\\ Actions\end{tabular}}} &
  \multicolumn{2}{c}{\textbf{\begin{tabular}[c]{@{}c@{}}Detected \\ Failures\end{tabular}}} \\ \cline{4-9} \cline{11-16} 
\multicolumn{2}{c}{\multirow{-1}{*}{\textbf{\begin{tabular}[c]{@{}c@{}}Compared  Approach\end{tabular}}}} &
  \multirow{-1}{*}{\textbf{WUT}} &
  $p$-value &
  $A_{12}$ &
  $p$-value &
  $A_{12}$ &
  $p$-value &
  $A_{12}$ &
  \multirow{-1}{*}{\textbf{WUT}} &
  $p$-value &
  $A_{12}$ &
  $p$-value &
  $A_{12}$ &
  $p$-value &
  $A_{12}$ \\ \midrule
\multicolumn{1}{c|}{\multirow{-2}{*}{\cellcolor[HTML]{F5F5F5}}} &
  \cellcolor[HTML]{F5F5F5}$\webexplor^5$ &
   &
  \cellcolor[HTML]{F5F5F5}\textbf{0.0022} &
  \cellcolor[HTML]{F5F5F5}\textbf{1.00} &
  \cellcolor[HTML]{F5F5F5}\textbf{0.0022} &
  \cellcolor[HTML]{F5F5F5}\textbf{1.00} &
  \cellcolor[HTML]{F5F5F5}0.0538 &
  \cellcolor[HTML]{F5F5F5}\textbf{0.85} &
   &
  \cellcolor[HTML]{F5F5F5}\textbf{0.0022} &
  \cellcolor[HTML]{F5F5F5}\textbf{1.00} &
  \cellcolor[HTML]{F5F5F5}\textbf{0.0050} &
  \cellcolor[HTML]{F5F5F5}\textbf{1.00} &
  \cellcolor[HTML]{F5F5F5}\textbf{0.0062} &
  \cellcolor[HTML]{F5F5F5}\textbf{0.99} \\
\multicolumn{1}{c|}{\multirow{-2}{*}{\cellcolor[HTML]{F5F5F5}SOTA}} &
  \cellcolor[HTML]{F5F5F5}$\marg^5$ &
   &
  \cellcolor[HTML]{F5F5F5}\textbf{0.0022} &
  \cellcolor[HTML]{F5F5F5}\textbf{1.00} &
  \cellcolor[HTML]{F5F5F5}\textbf{0.0022} &
  \cellcolor[HTML]{F5F5F5}\textbf{1.00} &
  \cellcolor[HTML]{F5F5F5}0.1994 &
  \cellcolor[HTML]{F5F5F5}\textbf{0.74} &
   &
  \cellcolor[HTML]{F5F5F5}1.0000 &
  \cellcolor[HTML]{F5F5F5}0.50 &
  \cellcolor[HTML]{F5F5F5}0.2265 &
  \cellcolor[HTML]{F5F5F5}\textbf{0.72} &
  \cellcolor[HTML]{F5F5F5}1.0000 &
  \cellcolor[HTML]{F5F5F5}0.49 \\
\multicolumn{1}{c|}{\multirow{-2}{*}{\cellcolor[HTML]{FFF8E1}}} &
  \cellcolor[HTML]{FFF8E1}$\tool^5$ \withoutcomm &
   &
  \cellcolor[HTML]{FFF8E1}\textbf{0.0022} &
  \cellcolor[HTML]{FFF8E1}\textbf{1.00} &
  \cellcolor[HTML]{FFF8E1}\textbf{0.0050} &
  \cellcolor[HTML]{FFF8E1}\textbf{1.00} &
  \cellcolor[HTML]{FFF8E1}0.2946 &
  \cellcolor[HTML]{FFF8E1}0.69 &
   &
  \cellcolor[HTML]{FFF8E1}0.0931 &
  \cellcolor[HTML]{FFF8E1}\textbf{0.81} &
  \cellcolor[HTML]{FFF8E1}1.0000 &
  \cellcolor[HTML]{FFF8E1}0.50 &
  \cellcolor[HTML]{FFF8E1}0.2946 &
  \cellcolor[HTML]{FFF8E1}0.69 \\
\multicolumn{1}{c|}{\multirow{-2}{*}{\cellcolor[HTML]{FFF8E1}Ablation}} &
  \cellcolor[HTML]{FFF8E1}$\tool^5$ \withoutqtran &
  \multirow{-4}{*}{EatingWell} &
  \cellcolor[HTML]{FFF8E1}\textbf{0.0152} &
  \cellcolor[HTML]{FFF8E1}\textbf{0.92} &
  \cellcolor[HTML]{FFF8E1}\textbf{0.0043} &
  \cellcolor[HTML]{FFF8E1}\textbf{0.97} &
  \cellcolor[HTML]{FFF8E1}0.7475 &
  \cellcolor[HTML]{FFF8E1}0.43 &
  \multirow{-4}{*}{GameSpot} &
  \cellcolor[HTML]{FFF8E1}0.5211 &
  \cellcolor[HTML]{FFF8E1}0.62 &
  \cellcolor[HTML]{FFF8E1}0.8099 &
  \cellcolor[HTML]{FFF8E1}0.44 &
  \cellcolor[HTML]{FFF8E1}0.5182 &
  \cellcolor[HTML]{FFF8E1}0.38 \\ \midrule
\multicolumn{1}{c|}{\multirow{-2}{*}{\cellcolor[HTML]{F5F5F5}}} &
  \cellcolor[HTML]{F5F5F5}$\webexplor^5$ &
   &
  \cellcolor[HTML]{F5F5F5}\textbf{0.0022} &
  \cellcolor[HTML]{F5F5F5}\textbf{1.00} &
  \cellcolor[HTML]{F5F5F5}\textbf{0.0022} &
  \cellcolor[HTML]{F5F5F5}\textbf{1.00} &
  \cellcolor[HTML]{F5F5F5}\textbf{0.0022} &
  \cellcolor[HTML]{F5F5F5}\textbf{1.00} &
   &
  \cellcolor[HTML]{F5F5F5}\textbf{0.0022} &
  \cellcolor[HTML]{F5F5F5}\textbf{1.00} &
  \cellcolor[HTML]{F5F5F5}\textbf{0.0022} &
  \cellcolor[HTML]{F5F5F5}\textbf{1.00} &
  \cellcolor[HTML]{F5F5F5}\textbf{0.0048} &
  \cellcolor[HTML]{F5F5F5}\textbf{1.00} \\
\multicolumn{1}{c|}{\multirow{-2}{*}{\cellcolor[HTML]{F5F5F5}SOTA}} &
  \cellcolor[HTML]{F5F5F5}$\marg^5$ &
   &
  \cellcolor[HTML]{F5F5F5}0.0931 &
  \cellcolor[HTML]{F5F5F5}\textbf{0.81} &
  \cellcolor[HTML]{F5F5F5}0.0931 &
  \cellcolor[HTML]{F5F5F5}\textbf{0.81} &
  \cellcolor[HTML]{F5F5F5}0.9372 &
  \cellcolor[HTML]{F5F5F5}0.53 &
   &
  \cellcolor[HTML]{F5F5F5}\textbf{0.0050} &
  \cellcolor[HTML]{F5F5F5}\textbf{1.00} &
  \cellcolor[HTML]{F5F5F5}\textbf{0.0022} &
  \cellcolor[HTML]{F5F5F5}\textbf{1.00} &
  \cellcolor[HTML]{F5F5F5}0.1978 &
  \cellcolor[HTML]{F5F5F5}\textbf{0.74} \\
\multicolumn{1}{c|}{\multirow{-2}{*}{\cellcolor[HTML]{FFF8E1}}} &
  \cellcolor[HTML]{FFF8E1}$\tool^5$ \withoutcomm &
   &
  \cellcolor[HTML]{FFF8E1}\textbf{0.0152} &
  \cellcolor[HTML]{FFF8E1}\textbf{0.92} &
  \cellcolor[HTML]{FFF8E1}0.0931 &
  \cellcolor[HTML]{FFF8E1}\textbf{0.81} &
  \cellcolor[HTML]{FFF8E1}0.9372 &
  \cellcolor[HTML]{FFF8E1}0.47 &
   &
  \cellcolor[HTML]{FFF8E1}\textbf{0.0411} &
  \cellcolor[HTML]{FFF8E1}\textbf{0.86} &
  \cellcolor[HTML]{FFF8E1}0.0649 &
  \cellcolor[HTML]{FFF8E1}\textbf{0.83} &
  \cellcolor[HTML]{FFF8E1}\textbf{0.0100} &
  \cellcolor[HTML]{FFF8E1}\textbf{0.96} \\
\multicolumn{1}{c|}{\multirow{-2}{*}{\cellcolor[HTML]{FFF8E1}Ablation}} &
  \cellcolor[HTML]{FFF8E1}$\tool^5$ \withoutqtran &
  \multirow{-4}{*}{ESPN} &
  \cellcolor[HTML]{FFF8E1}\textbf{0.0087} &
  \cellcolor[HTML]{FFF8E1}\textbf{0.94} &
  \cellcolor[HTML]{FFF8E1}\textbf{0.0022} &
  \cellcolor[HTML]{FFF8E1}\textbf{1.00} &
  \cellcolor[HTML]{FFF8E1}0.8726 &
  \cellcolor[HTML]{FFF8E1}0.54 &
  \multirow{-4}{*}{Gap} &
  \cellcolor[HTML]{FFF8E1}\textbf{0.0152} &
  \cellcolor[HTML]{FFF8E1}\textbf{0.92} &
  \cellcolor[HTML]{FFF8E1}\textbf{0.0260} &
  \cellcolor[HTML]{FFF8E1}\textbf{0.89} &
  \cellcolor[HTML]{FFF8E1}0.0530 &
  \cellcolor[HTML]{FFF8E1}\textbf{0.85} \\ \midrule
\multicolumn{1}{c|}{\multirow{-2}{*}{\cellcolor[HTML]{F5F5F5}}} &
  \cellcolor[HTML]{F5F5F5}$\webexplor^5$ &
   &
  \cellcolor[HTML]{F5F5F5}\textbf{0.0022} &
  \cellcolor[HTML]{F5F5F5}\textbf{1.00} &
  \cellcolor[HTML]{F5F5F5}\textbf{0.0022} &
  \cellcolor[HTML]{F5F5F5}\textbf{1.00} &
  \cellcolor[HTML]{F5F5F5}0.1727 &
  \cellcolor[HTML]{F5F5F5}\textbf{0.75} &
   &
  \cellcolor[HTML]{F5F5F5}\textbf{0.0050} &
  \cellcolor[HTML]{F5F5F5}\textbf{1.00} &
  \cellcolor[HTML]{F5F5F5}\textbf{0.0050} &
  \cellcolor[HTML]{F5F5F5}\textbf{1.00} &
  \cellcolor[HTML]{F5F5F5}\textbf{0.0050} &
  \cellcolor[HTML]{F5F5F5}\textbf{1.00} \\
\multicolumn{1}{c|}{\multirow{-2}{*}{\cellcolor[HTML]{F5F5F5}SOTA}} &
  \cellcolor[HTML]{F5F5F5}$\marg^5$ &
   &
  \cellcolor[HTML]{F5F5F5}\textbf{0.0152} &
  \cellcolor[HTML]{F5F5F5}\textbf{0.92} &
  \cellcolor[HTML]{F5F5F5}\textbf{0.0411} &
  \cellcolor[HTML]{F5F5F5}\textbf{0.86} &
  \cellcolor[HTML]{F5F5F5}0.1994 &
  \cellcolor[HTML]{F5F5F5}0.26 &
   &
  \cellcolor[HTML]{F5F5F5}\textbf{0.0022} &
  \cellcolor[HTML]{F5F5F5}\textbf{1.00} &
  \cellcolor[HTML]{F5F5F5}\textbf{0.0043} &
  \cellcolor[HTML]{F5F5F5}\textbf{0.97} &
  \cellcolor[HTML]{F5F5F5}1.0000 &
  \cellcolor[HTML]{F5F5F5}0.50 \\
\multicolumn{1}{c|}{\multirow{-2}{*}{\cellcolor[HTML]{FFF8E1}}} &
  \cellcolor[HTML]{FFF8E1}$\tool^5$ \withoutcomm &
   &
  \cellcolor[HTML]{FFF8E1}0.1797 &
  \cellcolor[HTML]{FFF8E1}\textbf{0.75} &
  \cellcolor[HTML]{FFF8E1}0.1320 &
  \cellcolor[HTML]{FFF8E1}\textbf{0.78} &
  \cellcolor[HTML]{FFF8E1}0.9372 &
  \cellcolor[HTML]{FFF8E1}0.47 &
   &
  \cellcolor[HTML]{FFF8E1}\textbf{0.0022} &
  \cellcolor[HTML]{FFF8E1}\textbf{1.00} &
  \cellcolor[HTML]{FFF8E1}0.0649 &
  \cellcolor[HTML]{FFF8E1}\textbf{0.83} &
  \cellcolor[HTML]{FFF8E1}0.2963 &
  \cellcolor[HTML]{FFF8E1}0.31 \\
\multicolumn{1}{c|}{\multirow{-2}{*}{\cellcolor[HTML]{FFF8E1}Ablation}} &
  \cellcolor[HTML]{FFF8E1}$\tool^5$ \withoutqtran &
  \multirow{-4}{*}{FoxNews} &
  \cellcolor[HTML]{FFF8E1}0.4225 &
  \cellcolor[HTML]{FFF8E1}0.65 &
  \cellcolor[HTML]{FFF8E1}0.1320 &
  \cellcolor[HTML]{FFF8E1}\textbf{0.78} &
  \cellcolor[HTML]{FFF8E1}0.8182 &
  \cellcolor[HTML]{FFF8E1}0.44 &
  \multirow{-4}{*}{GitHub} &
  \cellcolor[HTML]{FFF8E1}\textbf{0.0022} &
  \cellcolor[HTML]{FFF8E1}\textbf{1.00} &
  \cellcolor[HTML]{FFF8E1}0.2403 &
  \cellcolor[HTML]{FFF8E1}\textbf{0.72} &
  \cellcolor[HTML]{FFF8E1}0.4217 &
  \cellcolor[HTML]{FFF8E1}0.35 \\ \midrule
\multicolumn{1}{c|}{\multirow{-2}{*}{\cellcolor[HTML]{F5F5F5}}} &
  \cellcolor[HTML]{F5F5F5}$\webexplor^5$ &
   &
  \cellcolor[HTML]{F5F5F5}\textbf{0.0022} &
  \cellcolor[HTML]{F5F5F5}\textbf{1.00} &
  \cellcolor[HTML]{F5F5F5}\textbf{0.0022} &
  \cellcolor[HTML]{F5F5F5}\textbf{1.00} &
  \cellcolor[HTML]{F5F5F5}\textbf{0.0049} &
  \cellcolor[HTML]{F5F5F5}\textbf{1.00} &
   &
  \cellcolor[HTML]{F5F5F5}\textbf{0.0022} &
  \cellcolor[HTML]{F5F5F5}\textbf{1.00} &
  \cellcolor[HTML]{F5F5F5}\textbf{0.0050} &
  \cellcolor[HTML]{F5F5F5}\textbf{1.00} &
  \cellcolor[HTML]{F5F5F5}\textbf{0.0046} &
  \cellcolor[HTML]{F5F5F5}\textbf{1.00} \\
\multicolumn{1}{c|}{\multirow{-2}{*}{\cellcolor[HTML]{F5F5F5}SOTA}} &
  \cellcolor[HTML]{F5F5F5}$\marg^5$ &
   &
  \cellcolor[HTML]{F5F5F5}\textbf{0.0022} &
  \cellcolor[HTML]{F5F5F5}\textbf{1.00} &
  \cellcolor[HTML]{F5F5F5}\textbf{0.0087} &
  \cellcolor[HTML]{F5F5F5}\textbf{0.94} &
  \cellcolor[HTML]{F5F5F5}1.0000 &
  \cellcolor[HTML]{F5F5F5}0.50 &
   &
  \cellcolor[HTML]{F5F5F5}0.4848 &
  \cellcolor[HTML]{F5F5F5}0.36 &
  \cellcolor[HTML]{F5F5F5}0.2971 &
  \cellcolor[HTML]{F5F5F5}0.69 &
  \cellcolor[HTML]{F5F5F5}0.1275 &
  \cellcolor[HTML]{F5F5F5}\textbf{0.78} \\
\multicolumn{1}{c|}{\multirow{-2}{*}{\cellcolor[HTML]{FFF8E1}}} &
  \cellcolor[HTML]{FFF8E1}$\tool^5$ \withoutcomm &
   &
  \cellcolor[HTML]{FFF8E1}0.0931 &
  \cellcolor[HTML]{FFF8E1}\textbf{0.81} &
  \cellcolor[HTML]{FFF8E1}0.3095 &
  \cellcolor[HTML]{FFF8E1}0.69 &
  \cellcolor[HTML]{FFF8E1}0.6874 &
  \cellcolor[HTML]{FFF8E1}0.58 &
   &
  \cellcolor[HTML]{FFF8E1}0.3939 &
  \cellcolor[HTML]{FFF8E1}0.67 &
  \cellcolor[HTML]{FFF8E1}0.3095 &
  \cellcolor[HTML]{FFF8E1}0.69 &
  \cellcolor[HTML]{FFF8E1}\textbf{0.0301} &
  \cellcolor[HTML]{FFF8E1}\textbf{0.89} \\
\multicolumn{1}{c|}{\multirow{-2}{*}{\cellcolor[HTML]{FFF8E1}Ablation}} &
  \cellcolor[HTML]{FFF8E1}$\tool^5$ \withoutqtran &
  \multirow{-4}{*}{GameRant} &
  \cellcolor[HTML]{FFF8E1}\textbf{0.0260} &
  \cellcolor[HTML]{FFF8E1}\textbf{0.89} &
  \cellcolor[HTML]{FFF8E1}0.1320 &
  \cellcolor[HTML]{FFF8E1}\textbf{0.78} &
  \cellcolor[HTML]{FFF8E1}0.8095 &
  \cellcolor[HTML]{FFF8E1}0.56 &
  \multirow{-4}{*}{Smadex} &
  \cellcolor[HTML]{FFF8E1}0.8099 &
  \cellcolor[HTML]{FFF8E1}0.56 &
  \cellcolor[HTML]{FFF8E1}0.9372 &
  \cellcolor[HTML]{FFF8E1}0.53 &
  \cellcolor[HTML]{FFF8E1}0.2265 &
  \cellcolor[HTML]{FFF8E1}\textbf{0.72} \\ \bottomrule
\end{tabular}

}
\end{table*}

We conducted a statistical comparison between the results of \tool and those of baselines, as summarized in Table~\ref{tab:rq1_utest}.
Specifically, we performed the Mann-Whitney U test and evaluated the Vargha-Delaney $A_{12}$ effect size~\cite{arcuri2014hitchhiker, vargha2000critique}.
Here, a $p$-value $<0.05$ indicates a statistically significant difference between the results, while $A_{12}$ represents the probability that a randomly selected observation from \tool outperforms one from the compared approach, with $A_{12} > 0.71$ denoting a large effect~\cite{poulding2010efficient}. 
Results that satisfy these conditions are highlighted in bold. 

Table~\ref{tab:rq1_utest} shows that \tool consistently outperforms \webexplor across all WUTs in both explored states and executed unique actions.
Against \marg, \tool achieves statistically significant gains on five WUTs for explored states, with $A_{12}$ indicating stronger superiority on six WUTs.
For executed unique actions, it achieves significant improvements on five WUTs and stronger superiority on seven WUTs.
In general, \tool shows clear improvements over \webexplor and \marg, with statistically significant differences on most websites and $A_{12}$ values indicating strong superiority.

\begin{tcolorbox}
\textit{\textbf{Answer to RQ1}:}
{\tool} outperforms the state-of-the-art web testing approaches.
Specifically, \tool explores 33.3\% more states and executes 42.2\% more unique actions than \marg,  leading to the detection of more failures.
Statistical analysis further validates the significance of these improvements.
\end{tcolorbox}

\subsection{RQ2: Ablation Study}

\subsubsection{Method} 
We followed the same experimental settings and metrics as in RQ1 (Section~\ref{ssec:rq1_method}).
We compared \tool with two ablation variants (Section~\ref{sssec:ablation-variants}) to assess the performance gains from our key design innovations.

\subsubsection{Result}

As shown in Table~\ref{tab:rq1}, compared with the ablation variant in which multiple agents explore independently without communication, \tool achieves $27.1\%(=(978.8-769.9)/769.9)$ more explored states. 
Similarly, relative to the variant that removes QTRAN, \tool achieves $21.3\%(=(978.8-806.7)/806.7)$ more explored states. 
For action diversity, \tool executes 22.0\% more unique actions than the non-communication variant, and 21.3\% more than the non-QTRAN variant.
These advantages are particularly evident in highly interactive WUTs, such as Gap and GitHub.

Notably, both ablation variants employ DRL. 
Compared with \marg, which uses Q-learning, the non-communication variant achieves modest improvements on five out of the eight websites, with an overall increase of 4.9\% in terms of explored states.
This improvement highlights DRL’s advantage in large and complex web environments, where it learns rich state representations and enables more effective exploration than tabular methods like Q-learning.

Additionally, the non-QTRAN variant, which incorporates DRL into an experience-sharing scheme, achieves only limited improvement over the non-communication variant. 
This suggests that while communication remains important, the rules proposed by $\marg$ are insufficient to provide substantial benefits in DRL settings.
In contrast, \tool adopts the QTRAN algorithm, which is specifically designed to support the CTDE paradigm, allowing more effective cooperation between agents in high-dimensional state spaces.

\subsubsection{Statistical Analysis}

As shown in Table~\ref{tab:rq1_utest}, \tool explores significantly more states on four websites compared with the non-communication variant, and on five websites compared with the non-QTRAN variant, with $A_{12}$ values indicating large effect sizes on seven and five websites, respectively. 
In terms of executed unique actions, \tool achieves $A_{12}$ values indicating a large effect on five websites compared with the non-communication variant and on six websites compared with the non-QTRAN variant, meaning that \tool outperforms these variants in the majority of cases.

\begin{tcolorbox}
\textit{\textbf{Answer to RQ2}:}
{\tool} outperforms the ablation variants that focus on multi-agent communication.
\tool explores 27.1\% more states than the non-communication variant and 21.3\% more than the non-QTRAN variant.
The limited improvement of the latter variant over the former suggests that, while communication is important, \marg's experience-sharing rules are insufficient to provide substantial benefits in DRL settings.
\end{tcolorbox}

\subsection{RQ3: Coordination Overhead}

\subsubsection{Method}

Long-term performance reflects an approach's scalability in exploring large-scale WUTs, while also revealing the overhead associated with agent coordination algorithms as their policies (i.e., Q-tables or Q-networks) grow.
To assess this aspect, we conducted a long-term experiment on the website Gap, the most state-rich WUT identified in RQ1 (see Table~\ref{tab:rq1}), for a duration of 20 hours.
Both {\tool} and {\marg} were configured with five agents, and we recorded their \textit{hourly executed actions} to evaluate the degree of coordination overhead incurred during testing.
To mitigate randomness, we repeated the experiment six times.

\subsubsection{Result}

\begin{figure}[t]
    \centering
    \includegraphics[width=\columnwidth]{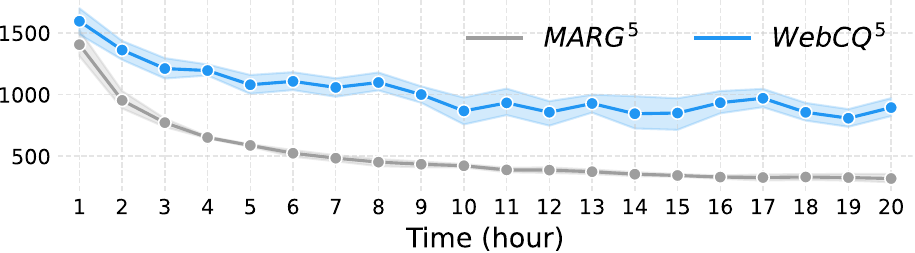}
    \vspace{-2em}
    \caption{Hourly executed actions (Y-axis) of \tool and \marg on Gap over 20 hours, where solid lines represent the mean values of six independent runs, and shaded areas represent their standard deviations.}
    \label{fig:rq2}
\end{figure}

\begin{figure*}[t]
    \centering
    \subcaptionbox{Explored States\label{sfig:state}}
      {\includegraphics[width=0.328\linewidth]{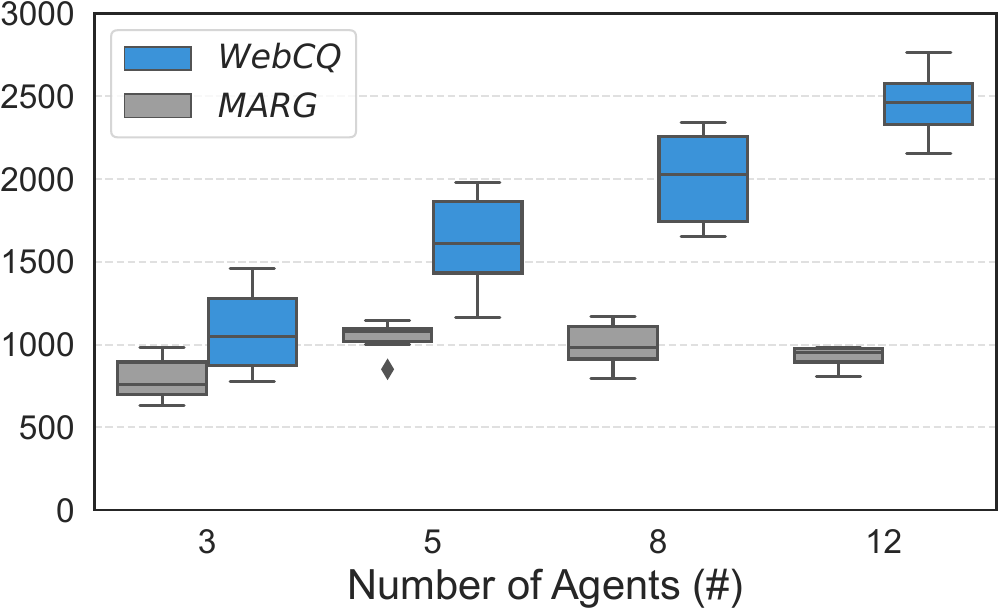}}
    \subcaptionbox{Executed Unique Actions\label{sfig:action}}
      {\includegraphics[width=0.328\linewidth]{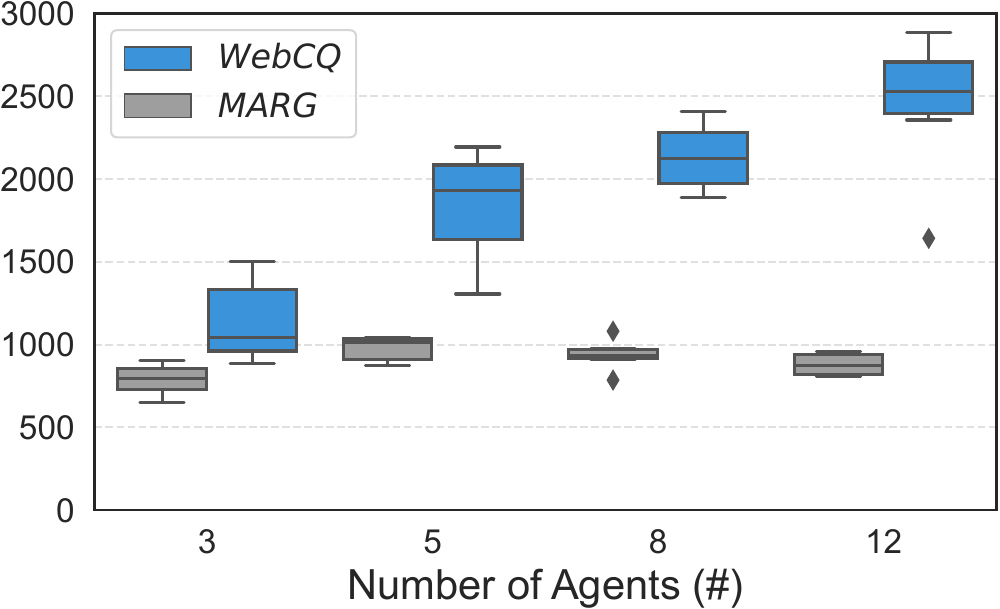}}
    \subcaptionbox{Detected Failures\label{sfig:failure}}
      {\includegraphics[width=0.328\linewidth]{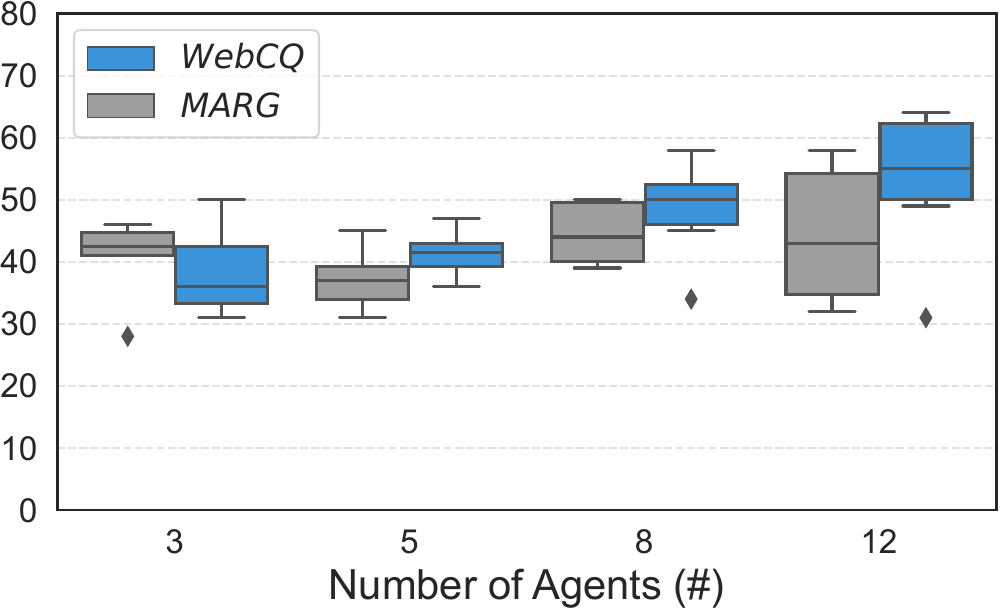}}  
    \vspace{-0.5em}
    \caption{Comparisons of \tool and \marg on Gap with increasing number of agents.}
    \label{fig:agent-num}
\end{figure*}

Figure~\ref{fig:rq2} presents the average numbers of hourly executed actions (and their standard deviations) of \tool and \marg on the Gap website over 20 hours.
At the beginning, both approaches exhibit a similar number of executed actions. 
This is because, in the initial phase, few states have been explored and the Q-table in \marg has not yet accumulated sufficient information, resulting in lower computational cost and faster coordination.
Meanwhile, \tool requires an additional initialization process for state and action embedding, resulting in a number of executed actions close to that of \marg during the first hour.
However, even at this early phase, \tool already outperforms \marg, suggesting that the DRL-based method achieves lower communication overhead than the tabular approach during the boosting stage.

As exploration progresses, a clear divergence emerges.
The number of executed actions by \marg declines rapidly and stabilizes at a low level, reflecting increasing coordination overhead as the Q-tables expand.
In contrast, although \tool also shows a downward trend, the decrease is more gradual and sometimes there are even rebounds, indicating sustained exploration capability.
Results from multiple runs further show that \marg produces stable but consistently low execution counts, while \tool exhibits higher variability yet maintains significantly higher counts in general.

These findings suggest that \tool consistently achieves better scalability and more efficient coordination in large and sparse web state spaces.
This advantage stems from the use of DQNs, which provide compact and generalizable representations beyond tabular methods, and from QTRAN, which further enhances multi-agent coordination through effective policy learning.

\begin{tcolorbox}
\textit{\textbf{Answer to RQ3}:}
\tool consistently outperforms \marg over a 20-hour period, showing better action execution efficiency.
This indicates that $\tool$ effectively mitigates the coordination overhead observed in \marg, thereby achieving better long-term performance.
\end{tcolorbox}

\subsection{RQ4: Effect of Agent Numbers}

\subsubsection{Method}

To evaluate the impact of agent numbers on the performance of MARL-based approaches, we ran \tool and \marg with varying agent numbers and compared their results.
The experiment was still conducted on Gap, for the same reason as in RQ3, with $N=3,5,8,12$ agents, consistent with {\marg}'s original settings \cite{fan2024can}.
Each configuration was run for three hours and was repeated six times.

\subsubsection{Result}

Figure~\ref{fig:agent-num} presents the results with varying agent numbers using box plots.
Across all agent configurations, $\tool$ consistently outperforms \marg on explored states and executed unique actions.
Although $\tool$ exhibits wider variability across runs, its median values are generally higher than those of \marg, reflecting better overall effectiveness.
This indicates that $\tool$ can effectively leverage additional agents to enhance decision-making, mitigating the impact of communication overhead.
An exception occurs at the maximum value of detected failures for $\tool^3$, which is slightly higher than that of $\tool^5$.
This could be attributed to the dynamic nature of the web environment and the stochasticity in action selection across runs.

In contrast, \marg shows diminishing improvements as the number of agents increases.
While increasing the number of agents from 3 to 5 leads to incremental gains, performance remains comparable or slightly declines at 8 agents, and drops when running 12 agents, especially in terms of explored states and executed unique actions.
Although running more agents may bring higher exploration potential, they also introduce greater communication overhead in a distributed Q-table architecture.
Compared to the commercial website used in their original study~\cite{fan2024can}, this trade-off was even more evident on Gap, suggesting that \marg’s tabular coordination strategy struggles to maintain efficiency when introducing more parallel agents.

These experimental results confirmed that $\tool$ achieves superior effectiveness and coordination scalability compared to \marg. 
Its DRL-based learning strategy, enhanced by QTRAN in a multi-agent scenario, enables efficient policy learning, making it more resilient to the overhead introduced by adding more agents.

\rev{Nevertheless, it does not imply that such overhead is completely eliminated.}
\rev{As shown in Table~\ref{tab:executed_actions}, the average number of actions executed per agent decreases from 985.22 in the 3-agent configuration to 576.24 in the 12-agent configuration, representing a reduction of 41.5\%.}
\rev{This decrease highlights the trade-off between greater parallelism and the additional coordination overhead incurred as the number of agents increases.}

\begin{table}[t]
\centering
\caption{\rev{Executed actions with different numbers of agents.}}
\label{tab:executed_actions}
\resizebox{0.75\columnwidth}{!}
{
\rev{
\begin{tabular}{@{}ccc@{}}
\toprule
\textbf{\shortstack{Number of\\Agents}} &
\textbf{\shortstack{Total Executed\\Actions}} &
\textbf{\shortstack{Executed Actions\\per Agent}} \\
\midrule
3  & 2,955.67 & 985.22 \\
5  & 4,317.00 & 863.40 \\
8  & 5,584.67 & 698.08 \\
12 & 6,914.83 & 576.24 \\
\bottomrule
\end{tabular}
}
}
\end{table}

\begin{tcolorbox}
\textit{\textbf{Answer to RQ4}:}
As the number of agents increases, the performance of {\tool} also improves steadily.
In contrast, the performance of {\marg} slightly decreases as more agents are involved.
QTRAN coordination makes \tool more resilient to the overhead introduced by adding more agents.
\end{tcolorbox}

\section{Threats to Validity}

Despite our best efforts to ensure the rigor of our study, several threats to validity still remain:

\textbf{The generalizability of our results may be limited, as experiments were conducted on only eight websites.}
MARL-based testing methods are particularly effective for large-scale websites, as they tend to achieve more significant performance gains on WUTs with rich content and interactive elements.
As a result, these methods are less suitable for open-source websites, which often have limited functionality.
To this end, we selected eight large-scale commercial websites and compared {\tool} with other MARL-based approaches.
To ensure representativeness, the selected websites span diverse categories and exhibit a substantial number of discoverable states, as shown in Table~\ref{tab:subjects} and Table~\ref{tab:rq1}.

\rev{\textbf{Given the large hyperparameter search space, the chosen hyperparameters may not be optimal across all websites.}
We conducted a pilot experiment on a representative website to identify an effective combination of parameter values and applied it to all experiments.
This strategy avoids subject-specific tuning and ensures a fair comparison across different subjects.
Nevertheless, further investigation of adaptive hyperparameter tuning may improve the performance of \tool under practical deployment scenarios.}

\textbf{Some degree of randomness was inevitable throughout our research.}
During \tool's working process, the DNN weights were randomly initialized, actions were selected using an $\varepsilon$-greedy strategy (Equation \ref{eq:epsilon-greedy}), and the mini-batch sampling strategy introduced additional variability in optimization directions.
To mitigate the impact of randomness, we repeated each experiment six times and performed statistical analyses.
This experimental setting is comparable to that of existing studies~\cite{zheng2021automatic,lan2024deeply,fan2024can,bauersfeld2014user}.

\section{Related Work}

RL-based techniques can optimize exploration strategies through continuous interactions with the environment~\cite{morales2020grokking}, making them promising for efficient GUI testing.
\textit{AutoBlackTest}~\cite{mariani2011autoblacktest} is an early-stage RL-based testing tool for desktop applications.
It uses the collection of GUI elements (i.e., element set) as the abstracted state and calculates the reward by GUI changes.
Bauersfeld et al.~\cite{bauersfeld2012reinforcement} adopted a similar approach, except they used the inverse of action frequencies as reward.
This reward design has also been adopted in later studies (e.g., \textit{WebExplor}~\cite{zheng2021automatic} or \textit{QExplore}~\cite{sherin2023qexplore}), referred to as ``curiosity''.
The exploration capabilities of such techniques can also be utilized in other tasks like bug reproduction~\cite{zhang2023automatically} and bug detection~\cite{guo2023effectively}.
Fan et al.~\cite{fan2023comprehensive} summarized Q-learning–based GUI testing techniques, highlighting effective configurations such as element-set–based states and curiosity-based rewards.
They also pointed out the single-agent limitation in testing large-scale websites.

DRL methods are instrumental in addressing the limitations of tabular RL algorithms.
They are particularly effective in managing the vast state spaces and dynamic behaviors of GUI applications.
Most of such research focuses on improving the exploration capabilities of testing agents on the Android platform, such as \textit{DeepGUIT}~\cite{collins2021deep}, \textit{ARES}~\cite{romdhana2022deep} and \textit{DinoDroid}~\cite{zhao2024dinodroid}.
Besides, \textit{PIRLTest} realizes platform-independent testing by embedding visual GUI states and employing a curiosity-driven exploration strategy to uncover untested application behaviors~\cite{yu2024effective}.
These DRL-based techniques demonstrate significant improvements over other methods in terms of code coverage and fault discovery capabilities.

MARL-based testing aims at coordinating multiple running RL agents, thereby mitigating the slow state exploration limitation of traditional RL-based testing techniques.
\textit{Fastbot}~\cite{cai2020fastbot} runs multiple Q-learning agents to collectively build navigation models for Android applications.
$GT^{PQL}$~\cite{mobilio2023gui} uses parallel Q-learning agents that collaborate and periodically synchronize their Q-models.
{\marg}~\cite{fan2024can} employs more efficient communication mechanisms among testing agents, but it remains constrained by traditional tabular RL.

\section{Conclusion \rev{and Future Work}}

We presented a novel deep MARL-based approach {\tool} for multi-agent web GUI testing.
{\tool} leverages QTRAN, a collaborative MARL algorithm, to coordinate multiple DRL agents and employs a lightweight synchronization method to adapt QTRAN to asynchronous web testing.
Each agent in \tool captures semantic and exploration-relevant information in its action vectors and combines them with the state vector to estimate Q-values through the Q-network.
Experiments demonstrate that {\tool} outperforms {\marg}, the state-of-the-art MARL-based approach, and two DRL-based ablation variants, and exhibits strong scalability.

\rev{Our future work includes extending \tool with other MARL algorithms, such as VDN~\cite{sunehag2017value} and QMIX~\cite{rashid2020monotonic}, to investigate the impact of different multi-agent learning architectures on web GUI testing.
Another interesting direction is to study failure reproduction in MARL-based testing environments, where parallel agent interactions and nondeterministic execution orders may introduce challenges in reliably reproducing detected failures.}

\begin{acks}
The authors sincerely appreciate all anonymous reviewers for their valuable comments, which have greatly improved this paper.
This work is supported by the National Natural Science Foundation of China (Grant No. 62372219).
\end{acks}

\section*{Data Availability Statement}


The evaluated artifact is available at~\cite{webcq_artifact}. Additional experimental data can be found at~\cite{webcq_data}.


\bibliographystyle{ACM-Reference-Format}

\let\WebCQOriginalBibitem\bibitem
\renewcommand{\bibitem}[2][]{%
  \def\WebCQCurrentBibKey{#2}%
  \def\WebCQColumnBreakBibKey{watkins1992q}%
  \ifx\WebCQCurrentBibKey\WebCQColumnBreakBibKey
    \newpage
  \fi
  \WebCQOriginalBibitem[#1]{#2}%
}

\bibliography{references}

\end{document}